\documentclass[useAMS,usenatbib]{mn2e}

\usepackage[T1]{fontenc}
\usepackage{pslatex}
\usepackage{amsmath}
\usepackage{amsfonts}
\usepackage{color}
\usepackage{url}
\usepackage{graphicx}
\usepackage{subfigure}
\usepackage{amssymb}
\usepackage{bm}
\usepackage{textcomp}
\usepackage{soul}
\usepackage{booktabs}
\usepackage{array}
\usepackage[draft]{hyperref}
 \graphicspath{{images/}}

{\newif\ifnotend
\notendtrue
\def\veclist{ABCDEFGHIJKLMNOPQRSTUVWXYZabcdefghijklmnopqrstuvwxyz.}
\def\top#1#2.{#1}
\def\tail#1#2.{#2.}
\loop\expandafter\xdef\csname v\expandafter\top\veclist\endcsname%
{{\noexpand\bf\expandafter\top\veclist}}
\edef\veclist{\expandafter\tail\veclist}
\if\veclist.\notendfalse\fi\ifnotend\repeat}

\def\e{{\rm e}}

\def\pa{\partial}

\mathchardef\mhyphen="2D

\title[A dynamical mechanism for the origin of nuclear rings]{A dynamical mechanism for the origin of nuclear rings}
\author[Sormani, Sobacchi, Fragkoudi, Ridley, Tre{\ss}, Glover, Klessen]{Mattia C. Sormani$^{1}$, Emanuele Sobacchi$^{2,3}$, Francesca Fragkoudi$^{4,5}$, Matthew Ridley$^6$, \newauthor Robin G. Tre{\ss}$^1$, 
Simon C.O. Glover$^1$, Ralf S. Klessen$^{1,7}$ \\
$^1$Universit\"{a}t Heidelberg, Zentrum f\"{u}r Astronomie, Institut f\"{u}r theoretische Astrophysik, Albert-Ueberle-Str. 2, 69120 Heidelberg, Germany \\
$^2$Physics Department, Ben-Gurion University, P.O.B. 653, Beer-Sheva 84105, Israel \\
$^3$Department of Natural Sciences, The Open University of Israel, 1 University Road, P.O.B. 808, Raanana 4353701, Israel \\
$^4$Max-Planck-Institut f\"{u}r Astrophysik, Karl-Schwarzschild-Stra{\ss}e 1, 85748, Garching bei M\"{u}nchen, Germany \\
$^5$GEPI, Observatoire de Paris, PSL Universit\'e, CNRS,  5 Place Jules Janssen, 92190 Meudon, France \\
$^6$Rudolf Peierls Centre for Theoretical Physics, 1 Keble Road, Oxford OX1 3NP, UK \\
$^7$Universit\"at Heidelberg, Interdiszipli\"ares Zentrum f\"ur Wissenschaftliches Rechnen, Im Neuenheimer Feld 205, 69120 Heidelberg, Germany
}

\begin{document}
\hyphenation{kruijs-sen}

\date{}

\def\p{\partial}
\def\Omegap{\Omega_{\rm p}}

\newcommand{\di}{\mathrm{d}}
\newcommand{\bfx}{\mathbf{x}}
\newcommand{\bfe}{\mathbf{e}}
\newcommand{\vlos}{\mathrm{v}_{\rm los}}
\newcommand{\Tspin}{T_{\rm s}}
\newcommand{\Tb}{T_{\rm b}}
\newcommand{\degree}{\ensuremath{^\circ}}
\newcommand{\Th}{T_{\rm h}}
\newcommand{\Tc}{T_{\rm c}}
\newcommand{\bfr}{\mathbf{r}}
\newcommand{\bfv}{\mathbf{v}}
\newcommand{\pc}{\,{\rm pc}}
\newcommand{\kpc}{\,{\rm kpc}}
\newcommand{\Myr}{\,{\rm Myr}}
\newcommand{\Gyr}{\,{\rm Gyr}}
\newcommand{\kms}{\,{\rm km\, s^{-1}}}
\newcommand{\de}[2]{\frac{\partial #1}{\partial {#2}}}
\newcommand{\cs}{c_{\rm s}}
\newcommand{\rb}{r_{\rm b}}
\newcommand{\rqu}{r_{\rm q}}
\newcommand{\nuP}{\nu_{\rm P}}
\newcommand{\thetaobs}{\theta_{\rm obs}}
\newcommand{\hatn}{\hat{\textbf{n}}}
\newcommand{\hatt}{\hat{\textbf{t}}}
\newcommand{\hatx}{\hat{\textbf{x}}}
\newcommand{\haty}{\hat{\textbf{y}}}
\newcommand{\hatz}{\hat{\textbf{z}}}
\newcommand{\hatX}{\hat{\textbf{X}}}
\newcommand{\hatY}{\hat{\textbf{Y}}}
\newcommand{\hatZ}{\hat{\textbf{Z}}}
\newcommand{\hatN}{\hat{\textbf{N}}}

\maketitle

\begin{abstract}
We develop a dynamical theory for the origin of nuclear rings in barred galaxies. In analogy with the standard theory of accretion discs, our theory is based on shear viscous forces among nested annuli of gas. However, the fact that gas follows non circular orbits in an external barred potential has profound consequences: it creates a region of reverse shear in which it is energetically favourable to form a stable ring which does not spread despite dissipation. Our theory allows us to approximately predict the size of the ring given the underlying gravitational potential. The size of the ring is loosely related to the location of the Inner Lindblad Resonance in the epicyclic approximation, but the predicted location is more accurate and is also valid for strongly barred potentials. By comparing analytical predictions with the results of hydrodynamical simulations, we find that our theory provides a viable mechanism for ring formation if the effective sound speed of the gas is low ($\cs\lesssim1\kms$), but that nuclear spirals/shocks created by pressure destroy the ring when the sound speed is high ($\cs\simeq10\kms$). We conclude that whether this mechanism for ring formation is relevant for real galaxies ultimately depends on the effective equation of state of the ISM. Promising confirmation comes from simulations in which the ISM is modelled using state-of-the-art cooling functions coupled to live chemical networks, but more tests are needed regarding the role of turbulence driven by stellar feedback. If the mechanism is relevant in real galaxies, it could provide a powerful tool to constrain the gravitational potential, in particular the bar pattern speed.
\end{abstract}


\section{Introduction}
\label{sec:intro}

Rings of gas and stars are a common morphological feature of galaxies \citep[e.g.][]{ButaCombes1996,Knapen2005,Comeron2010,Comeron2013,Buta2017a,Buta2017b}. They are most often associated with a stellar bar and in the parlance of galaxy morphology are classified in three main types, primarily according to their size: nuclear rings, inner rings and outer rings. 

Nuclear rings are the smallest type of ring and are commonly found at the centre of barred galaxies. They are usually sites of intense star formation. The radii of nuclear rings range from a few tens of parsec to $\sim$3 kpc \citep[e.g.][]{Knapen2005,Comeron2010,Comeron2013}. Our Galaxy is likely to have such a ring with radius $\simeq 150$ pc, which corresponds to the Central Molecular Zone \citep[e.g.][]{Liszt2009,Molinari+2011,Kruijssen+2015,Henshaw+16a,Sormani2018}. Nuclear rings are a gas reservoir for the accretion disc that surrounds the super-massive black hole that is present at the centre of most galaxies, although it is not clear how the gas migrates from the nuclear ring ($r \sim 100$pc) down to the accretion discs at much smaller radii ($r \sim 10^{-3}$pc) (see for example \citealt{Phinney1994,Combes2001} and also \citealt{Li+2017}).

It is currently unclear what exactly determines the size of nuclear rings \citep[e.g.][]{VandevenChang2009,Comeron2010}. It is clear that there must be some connection between the size of the ring and the Inner Lindblad Resonance (ILR) of the underlying potential \citep[e.g.][]{ButaCombes1996}, and that gas in the ring must flow on a family of non-circular orbits mildly elongated perpendicular to the bar called $x_2$ orbits \citep[e.g.][]{ReganTeuben2003,Maciejewski2004b,Kim++2012a,SBM2015a}. These two facts are not contradictory, since the size and extent of the $x_2$ orbital family is correlated to the location of the ILR(s) \citep[e.g.][]{contopoulosreview,Athan92a}. Virtually all modern explanations for the formation of rings broadly agree with these two statements, but they differ in the details of how the associations are made.

The most widely accepted theory, which we shall call resonant theory, suggests that the reason why nuclear rings are associated with the ILR is that a spiral pattern is induced by the bar such that it experiences net torques from the barred potential that change sign at each resonance \citep[e.g.][]{Combes1988,Combes1996b,ButaCombes1996}. Thus, as \cite{ButaCombes1996} describe it, `rings are formed by gas accumulation at the Lindblad resonances, under the continuous action of gravity torques from the bar pattern'. However, the concept of the ILR is well defined only for a weak bar. One may extend it to the case of a strong bar based on the $x_2$ orbital family \citep[see for example the discussion in section 3.2 of][]{Athan92a}, but then the $x_2$ orbits are in a sense all `resonant', and the resonant theory is unable to predict which orbits will be populated by gas. 

The sound speed of the gas $\cs$ is an important parameter of the flow. By running two-dimensional hydrodynamical simulations in an externally imposed barred potential, \cite{Kim++2012a} (see also \citealt{PatsisAthanassoula2000}) find that the size and morphology of the nuclear region depends on $\cs$, even if the the underlying gravitational potential and therefore the ILR position stays exactly the same. It is not clear how this finding can be reconciled with the resonant theory.

There is also evidence that viscosity plays a key role in the formation of rings. This is indicated by the fact that simulations using grid codes, which model the gas as a continuous isothermal fluid and typically do not include explicit viscous terms, can produce a variety of morphologies ranging from nuclear rings to nuclear spirals \citep[e.g.][]{Maciejewski2004b,Kim++2012a,Li+2015,SBM2015a,SBM2015b,Fragkoudi+2017}, while simulations that use the sticky-particle technique, which models the gas as inelastically colliding test particles and therefore contain some explicit amount of dissipation, almost invariably produce nuclear rings, and almost never nuclear spirals \citep[e.g.][]{Schwarz1981,CombesGerin1985,Byrd+1994,JenkinsBinney94,RautiainenSalo2000,Rautiainen+2002,Rautiainen+2004,RFC2008}. Thus viscosity seems to favour the development of rings, and how this can be reconciled with the resonant theory is also unclear.\footnote{According to the resonant theory, the key to ring formation is the generation of a spiral pattern on which gravity torques can act. Since as we discuss below in Sect. \ref{sec:results2} (see also \citealt{SBM2015b}) viscosity does not play a major role in the generation of the spiral pattern, according to the resonant theory viscosity should play little role in the formation of rings. Indeed, the spiral pattern is generated by pressure, not viscosity. This poses another puzzling question: if the idea underlying the resonant theory is correct, we would expect spirals created by pressure to always eventually develop into rings because of gravity torques. Yet, simulations show that this often does not happen, with the gas maintaining a quasi-steady nuclear spiral pattern that does not develop into rings \citep[e.g.][]{Kim++2012a,Li+2015}.}

Hence, two key parameters are the sound speed of the gas and the amount of viscosity. In order to understand the formation of rings we need to understand exactly what role they play. Unfortunately, to the best of our knowledge there are no studies that explore the effects of viscosity on gas flowing on non-circular orbits in a non-axisymmetric galactic potential,\footnote{A similar problem has  been studied extensively in the context of planetary ring dynamics, see Sect. \ref{sec:planetary}.} with the exception of the sticky-particles simulations that, however, are difficult to analyse because the viscosity they include is not the same as the classic Navier-Stokes viscosity.

In this paper, we aim to study the effects of viscosity and pressure on a disc of gas flowing on non-circular orbits in an externally imposed barred potential. We find that viscosity has consequences that can be significantly different from the circular case, which may be related to the origin of nuclear rings. We also aim to understand the interplay between viscosity and pressure, the latter being the other key parameter of the flow. Finally, we want to investigate whether the ring size can be predicted more precisely, i.e. beyond establishing a loose connection between the ring size and the location of the ILR.

The paper is structured as follows. In Section \ref{sec:accretion} we discuss a simple effect based on the viscous theory of accretion discs that is apparently only of academic interest, but will turn out to be a useful toy model to understand the mechanism discussed in Section \ref{sec:theory}. In Section \ref{sec:theory} we discuss the key mechanism that forms the basis of our picture for the formation of nuclear rings. We show that viscosity can create a trapping region where it is energetically favourable to form a ring. However, in common with viscous accretion disc theory in its simplest form, throughout this section we make one critical assumption: we neglect pressure. Hence in Section \ref{sec:resultsvisc} and \ref{sec:results2} we compare the analytical predictions of Section \ref{sec:theory} with the results of idealised isothermal simulations of a disc of gas flowing in an externally imposed barred potential that include various amounts of explicit viscosity and pressure. In Section \ref{sec:chemresults} we compare with a much more realistic simulation of the entire bar region of the Milky Way that models the ISM using state-of-the-art cooling functions and chemical networks rather than a simple isothermal approximation. In Section \ref{sec:discussion} we discuss (i) our results and approximations; (ii) what have we learned about the effects of viscosity and pressure; (iii) relation to other works, and (iv) relevance for real galaxies and possible applications. Section \ref{sec:summary} sums up.

\section{A consideration from the elementary theory of accretion discs} \label{sec:accretion}

\begin{figure}
\centering
\includegraphics[width=0.48\textwidth]{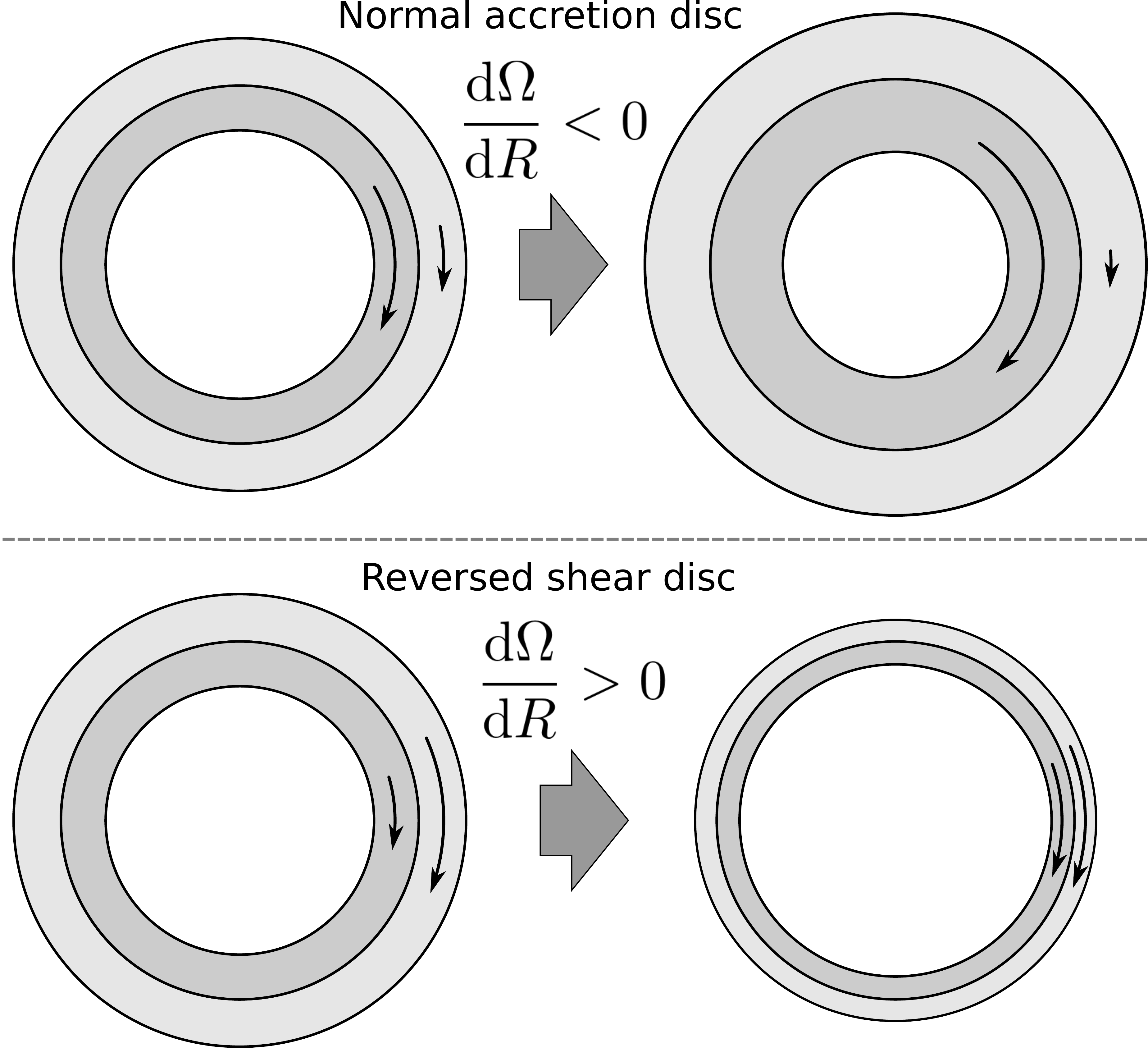}
\caption{When $\mathrm{d}\Omega/\mathrm{d}R<0$, as in a Kepler potential, the elementary theory of viscous accretion discs tells us that a ring of gas spreads in time. In a hypothetical situation where $\mathrm{d}\Omega/\mathrm{d}R>0$, the ring would tighten in time. There is no accretion in this case and one gets a ring-forming disc.}
\label{fig:accretion}
\end{figure}

In the elementary theory of accretion discs we learn that, as a consequence of viscous torques between adjacent annuli of gas, mass flows inwards and angular momentum flows outwards \citep[e.g.][]{ShakuraSunyaev1973,LyndenBellPringle1974,Pringle1981}. This situation is schematically depicted in the top panel of Fig. \ref{fig:accretion}: the inner ring rotates \emph{faster} than the outer ring, thus friction between the two will try to slow down the inner ring and speed up the outer ring. Angular momentum is transferred from the inner ring to the outer ring. So, if the inner ring loses angular momentum, but is forced to remain on a circular orbit, it must move inward to the radius corresponding to its new angular momentum, which increases its circular velocity.\footnote{This is the `donkey effect', see \cite{LyndenBellKalnajs1972} or Box 3.3 in \cite{BT2008}.} The outer ring moves outward, decreasing its circular velocity. The net result is that the ring spreads. The equations for the time evolution of the surface density $\Sigma(R,t)$ have much in common with the heat (diffusion) equation.

The previous result relies on the fundamental assumption that $\mathrm{d} \Omega/\mathrm{d} R<0$, i.e. that the angular velocity $\Omega(R)$ is a decreasing function of radius. This is true for the Kepler potential and in all practical cases in galaxies. However, let us ignore this for a moment and suppose that $\mathrm{d} \Omega/\mathrm{d} R>0$. What would happen in this case? The situation is depicted in the lower panel of Fig. \ref{fig:accretion}. The inner ring rotates \emph{slower} than the outer ring, thus friction between the two will try to speed up the inner ring and slow down the outer ring. Angular momentum is transferred from the outer ring to the inner ring. So, if everything is forced to remain on a circular orbit, the inner ring must move outwards, and the outer ring moves inwards. Therefore, we reach the conclusion that \emph{if $\mathrm{d}\Omega/\mathrm{d}R>0$, a ring of finite width becomes tighter and tighter}. There is no accretion in this case - instead, a disc with some density fluctuations in the initial surface density will break up into rings. The equations for the time evolution of the surface density $\Sigma(R,t)$ would look like the heat equation, but with heat flowing from cold to hot!\footnote{Indeed, the analysis of \cite{LyndenBellPringle1974} -- that shows that the most energetically favourable situation is reached when most of the gas is driven to the centre while all angular momentum has been transported to infinity by an infinitesimal amount of mass -- is not valid in the case $\mathrm{d}\Omega/\mathrm{d}R>0$ (see their last equation on page 606 and related remarks).}

These considerations seem only of academic interest, since $\mathrm{d}\Omega/\mathrm{d}R>0$ never actually happens in the rotation curves of real galaxies. However, we will see that this `reversed shear' disc serves as a useful toy model to understand what actually happens when gas follows non-circular orbits in a barred potential. The same idea has also been successfully applied to explain the confinement and narrow edges of rings in the context of planetary ring dynamics (see Sect. \ref{sec:planetary}).

\section{A mechanism for the formation of nuclear rings} \label{sec:theory}

Our starting point is the well known fact that gas flowing in the central regions of a barred potential does not follow circular orbits, but a family of closed orbits elongated perpendicular to the bar called $x_2$ orbits \citep[e.g.][]{contopoulosreview,SBM2015a}.

\begin{figure}
\centering
\includegraphics[width=0.48\textwidth]{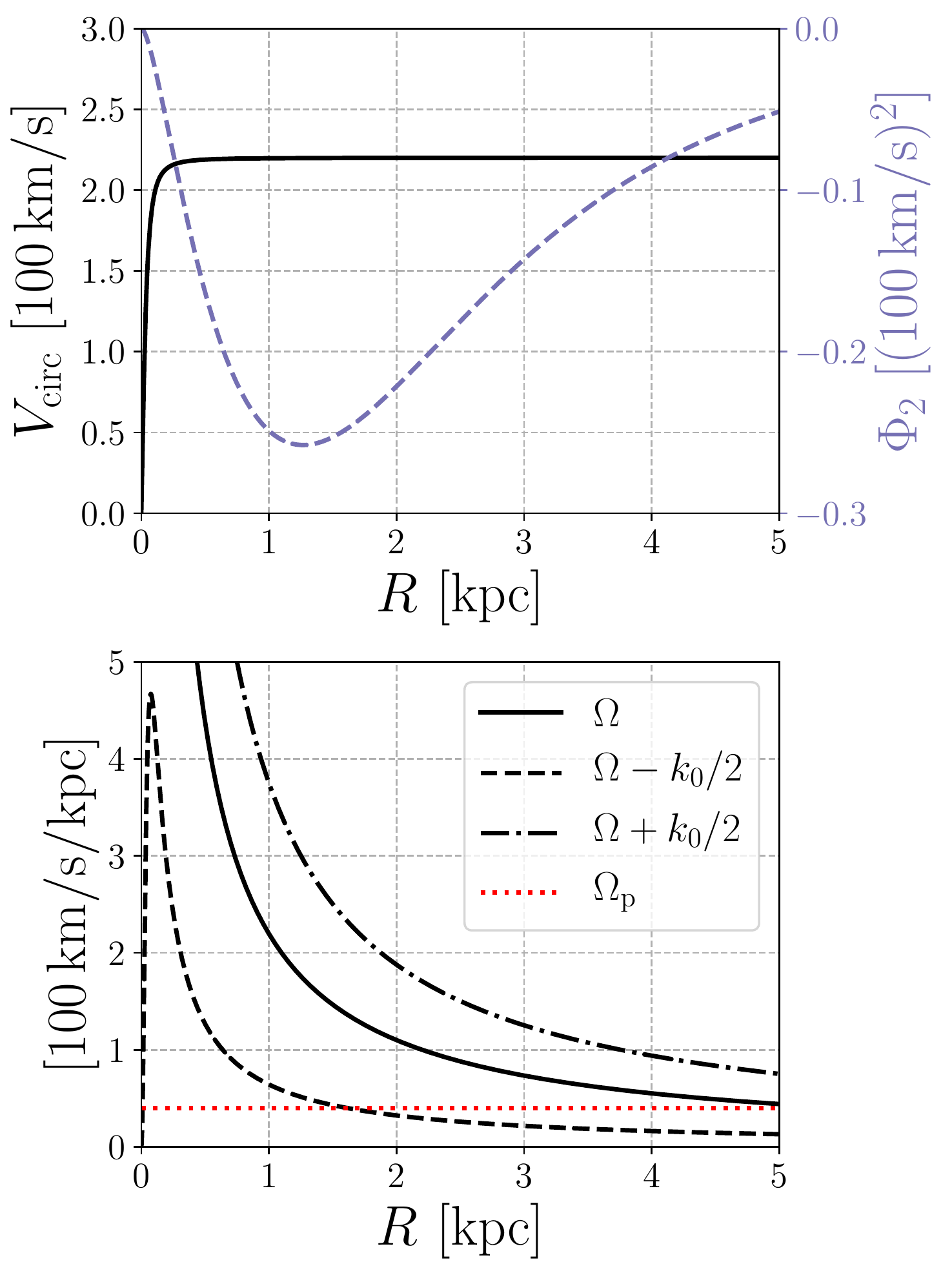}
\caption{\emph{Top panel}: the circular velocity curve $v_{\rm circ}(R) = \sqrt{R \di \Phi_0/\di R}$ (full black line) and the quadrupole $\Phi_2(R)$ (dashed line) of the potential described in Section \ref{sec:theory}. \emph{Bottom panel}: the curves $\Omega$ and $\Omega \pm k_0/2$, where $k_0^2= (2 \Omega/ R) \di (\Omega R^2)/ \di R$ is the epicyclic frequency. The red dashed line indicates the value of the bar pattern speed. The ILR is located at the intersection between the horizontal $\Omegap$ line and the $\Omega-k_0/2$ curve.}
\label{fig:SBM_2}
\end{figure}

As a concrete example, let us consider a simple analytical barred potential of the following form:
\begin{equation} \label{eq:sbmphi}
\Phi(R,\theta) = \Phi_0(R) + \Phi_2(R) \cos(2\theta),
\end{equation}
where $R,\theta$ are standard polar coordinates. For the axisymmetric part we take a simple logarithmic potential:
\begin{equation} \label{eq:log}
\Phi_0(R) = \frac{v_0^2}{2}  \log(R^2 + R_c^2),
\end{equation}
where $v_0=220 \kms$ and $R_c = 0.05 \kpc$. For the quadrupole $\Phi_2$ we adopt the same function that \cite{SBM2015c} used to simulate the gas flow in model Milky Way galaxies, which is described in detail in Appendix \ref{appendix:quadrupole}. The potential rigidly rotates with a constant pattern speed of $\Omegap = 40 \kms \kpc^{-1}$. Fig. \ref{fig:SBM_2} plots the circular velocity curve, the quadrupole and the resonance diagram in the epicyclic approximation.

\begin{figure*}
\centering
\includegraphics[width=1.0\textwidth]{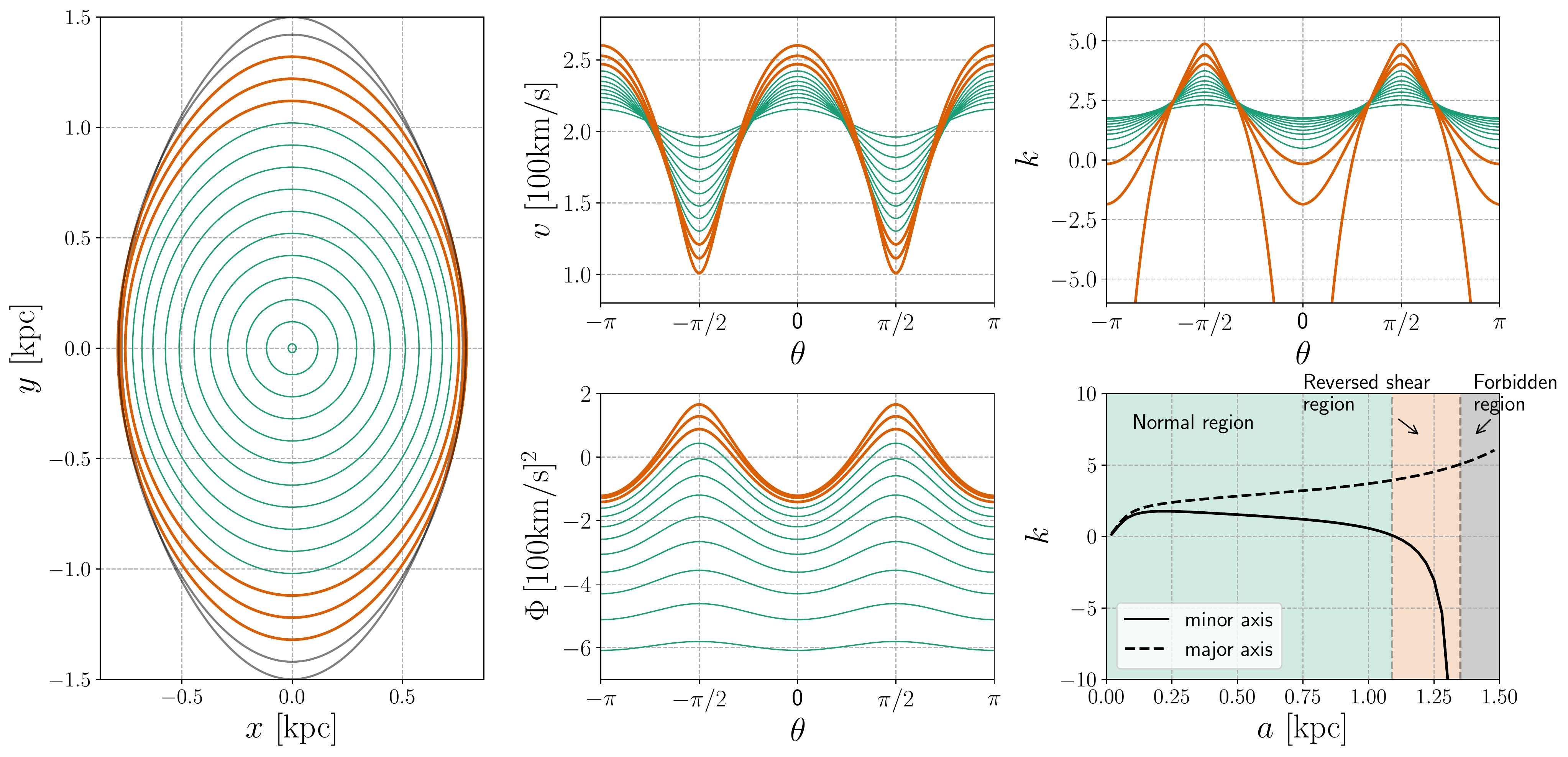}
\caption{\emph{Left:} $x_2$ orbits calculated in the barred potential described in Section \ref{sec:theory}. The bar major axis is horizontal. Green indicates orbits in which shear is always as in a `normal' accretion disc, while orange indicates orbits in which portions of `reversed shear' are present. Grey indicates `forbidden' orbits (see text). \emph{Top-middle:} modulus of the velocity calculated in a frame rotating at $\Omega_{\rm p}$ for the same orbits shown in the left panel. \emph{Bottom-middle:} gravitational potential along the orbits. \emph{Top-right:} the value of $k$ along orbits, i.e. the torque per unit length and per unit surface density that is exerted on orbit contours (see Eq. \ref{eq:k}). Units are normalised so that $\nu=1$, and are $[k]=100 \kms$. \emph{Bottom-right:} values of $k$ along the minor and major axis as a function of the semi-major axis of the orbits. When the full line is negative, a portion of the orbit with reversed shear exists.}
\label{fig:SBM_1}
\end{figure*}

The left panel of Fig. \ref{fig:SBM_1} shows $x_2$ orbits for this potential. These illustrate some fairly general characteristics of the $x_2$ orbital family. Orbits become more and more elongated as we move to larger values of the semi-major axis $a$. The $x_2$ family has a finite radial extent, and there are no $x_2$ orbits beyond a maximum value of the semi-major axis. This puts an obvious upper limit on the maximum size of the ring, i.e. it must be less than or equal to the size of the largest $x_2$ orbit, which for this particular potential is $a_{\rm max} \simeq 1.5 \kpc$. Note that the ILR is at $R_{\rm ILR} \simeq 1.6 \kpc$ (see intersection of $\Omega - k_0/2$ curve and $\Omegap$ in Fig. \ref{fig:SBM_2}).

\begin{figure}
\centering
\includegraphics[height=0.3\textwidth]{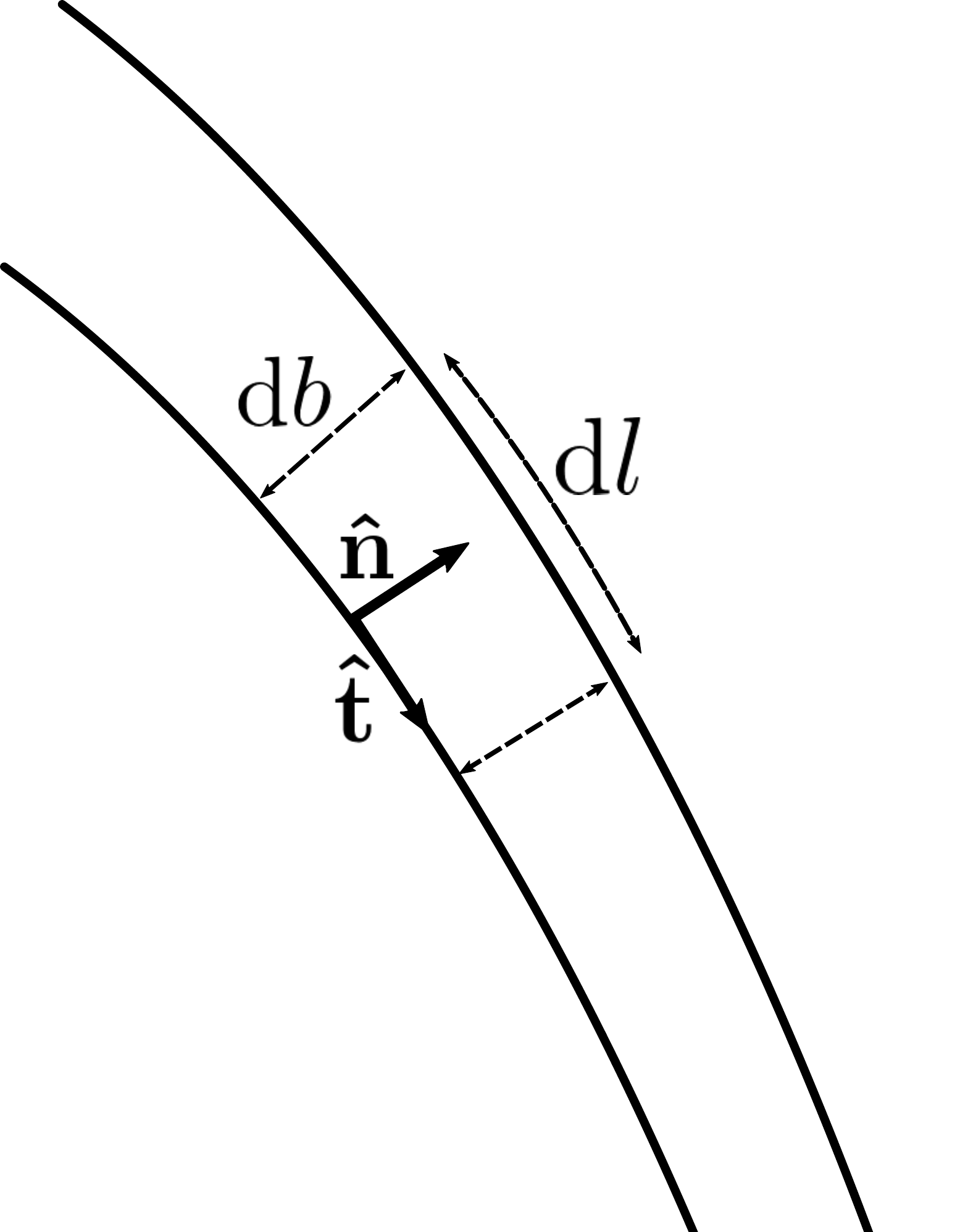}
\caption{Schematic diagram showing a portion of two adjacent orbits.}
\label{fig:diagram}
\end{figure}

Imagine constructing a steady state configuration of gas flowing on such orbits. To do that, consider a disc of gas comprised by nested annuli, each annulus covering the region between an $x_2$ orbit and an adjacent one. Ignoring pressure and viscous forces, a steady state can be constructed by taking the surface density in each annulus to be inversely proportional to the speed along the orbit $v$ and the width of the annulus $\di b$, that is
\begin{equation} \label{eq:rhoorbit}
\rho_{\rm annulus} = \frac{\alpha}{v \di b},
\end{equation}
where $\alpha$ is a constant that can be chosen independently for each annulus, and determines the mass of the annulus. 

Note that this construction is possible only for orbits with semi-major axis $a \lesssim 1.35 \kpc$ (green and orange in Fig. \ref{fig:SBM_1}): orbits with with semi-major axis greater than this value (shown in grey) cross adjacent orbits, so nested annuli cannot be constructed. Therefore, gas on these orbits would shock and plunge onto inner orbits. We shall call orbits in the grey area the `forbidden orbits'.

In the absence of pressure and viscous forces, steady states constructed this way will survive forever. What happens if we introduce some small viscosity, so that adjacent annuli can exert torques on each other, like they do in accretion discs? Let us calculate the torque that an annulus exerts on an adjacent one. Viscosity creates forces between layers of fluid that move relative to each other. The viscous force per unit length and unit surface density acting on an orbit contour can be written as:
\begin{equation}
f_i = n_j \sigma_{ij}
\end{equation}
where $\hatn$ is the normal to the contour (see Fig. \ref{fig:diagram}) and 
\begin{equation} \label{eq:sigmaij}
\sigma_{ij} = \nu \left( \pa_i v_j + \pa_j v_i - \frac{2}{3} \delta_{ij}\left( \nabla \cdot \bfv \right) \right)
\end{equation}
is the viscous stress tensor, where $\nu$ is the coefficient of kinematic viscosity with dimensions $[\nu] = \rm length^2 \times time^{-1}$.  The torque is the cross product of the radius and the force. Therefore, the torque per unit length and unit surface density applied to the contour is
\begin{equation} \label{eq:k}
\mathbf{k} = \bfr \times \mathbf{f},
\end{equation}
where $\bfr$ is the radial vector from the centre, and since our problem is two-dimensional we can write $\mathbf{k} = k\ \hatz$. Equation \eqref{eq:k} gives the torque on a \emph{contour}. It can be seen as the flux of angular momentum through the contour. To find the \emph{net} torque acting over a region bounded by a closed contour we need to integrate Eq. \eqref{eq:k} over the contour. 

For circular orbits Eq. \eqref{eq:k} reduces to
\begin{equation} \label{eq:kcirc}
k_{\rm circ} = - \nu R^2 \left( \frac{\di \Omega}{\di R} \right)
\end{equation}
thus we see that regions of `normal shear' ($\di \Omega/{\di R}<0$), in which angular momentum is flowing outwards, correspond to $k_{\rm circ}>0$, while regions of `reversed shear' ($\di \Omega/{\di R}>0$), in which angular momentum is flowing inwards, correspond to $k_{\rm circ}<0$. This correspondence between the sign of $k$ and the direction of the angular momentum flux remains true also in the non-circular case. Thus, regions of $k<0$ are regions of `reversed shear'.

The top-right panel in Fig. \ref{fig:SBM_1} plots the quantity $k$ along the $x_2$ orbits as a function of the polar angle $\theta$. For orbits in the green region, $k>0$ everywhere. This region behaves like a `normal' accretion disc with $\di \Omega/\di R<0$. Orange orbits instead have portions around $\theta=0$ and $\theta=\pi$ (which correspond to the $x$ axis) where $k<0$. These are portions of reversed shear where \emph{an inner orbit tries to slow down an outer orbit}, as in the circular case with $\di \Omega / \di R>0$. The reversed shear portions become larger as we move to outer orbits, and eventually the value of $k$ diverges on the $x$ axis as we enter the `forbidden region' (grey orbits). This suggests that some orbits within the orange region will be unstable for the formation of rings, in analogy with the considerations in Section \ref{sec:accretion}. A ring of gas flowing on these orbits would survive without spreading in time.

Intuitively, the reversal as we go from the $y$ to the $x$ axis happens for the following reason. Consider two adjacent orbits in the orange region. At the point where they intercept the $y$ axis, the angular velocity of the outer one is smaller than the angular velocity of the inner one, as in a normal accretion disc. However, because the elongation of the orbits increases as we move outwards, the outer orbit reduces its distance to the centre more than the inner orbit does as they approach the $x$ axis. The greater loss of gravitational potential energy of the outer orbit boosts its angular velocity and makes it \emph{greater} than the angular velocity of the inner orbit as they cross the $x$ axis, thus effectively producing a viscous force which tries to increase the velocity of the inner orbit, mimicking what would happen in the circular case for $\di \Omega / \di R>0$.

The above considerations lead to the following conclusions:
\begin{enumerate}
\item The fact that $x_2$ orbits are non-circular implies that there are regions of `reversed shear' (orange in Fig. \ref{fig:SBM_1}).
\item It is likely that this leads to the existence of regions which are unstable for the formation of rings and where a ring could survive without spreading in time despite dissipation.
\end{enumerate}

\subsection{Predicting the size of the ring} \label{sec:predictions}

In the last section we have argued that regions which are unstable for the formation of rings must be contained within the `reversed shear' region (orange in Fig. \ref{fig:SBM_1}), which is defined as the region where \emph{some} reverse shear is present along $x_2$ orbits. This gives a possible range where a ring may form. However, it is unlikely that rings can form throughout the whole orange region, because the `reverse shear' needs not only to be present, but also to be dominant. Thus, it is more likely that rings will form only in a sub-region within the orange region. Can we be more precise and narrow down this sub-region, so as to predict more precisely where a ring should form? 

The considerations of the previous section only apply locally along an orbit. But along the same $x_2$ orbit there can be both portions of normal and reversed shear. To determine the average behaviour of the gas and whether the net transport of angular momentum is directed outwards or inwards, one needs to integrate $\mathbf{k}$ along a full orbit contour, i.e. to calculate
\begin{equation} \label{eq:K}
\mathbf{K} = \int \rho \mathbf{k} \ \di l \;,
\end{equation}
where $\di l$ is the length of the element of the contour (see Fig \ref{fig:diagram}). Since we are in two dimensions we can write $\mathbf{K} = K \hatz$. The quantity $K$ can be thought of as the flux of angular momentum through the contour defined by an entire orbit. We use the convention that $K>0$ means angular momentum flux is directed outwards. Therefore, the difference $\delta K~=~K(a)~-~K(a + \di a) $ gives the net angular momentum gained by the annulus comprised by the orbits with semi-major axis $a$ and $a+\di a$. When $K$ is increasing (decreasing), the annulus loses (gains) angular momentum. Therefore, where $K$ is increasing (decreasing) matter moves inwards (outwards). Hence, the curve $K(a)$ potentially contains interesting information about locations where a ring could form. However, as we shall now see, to complicate things is the fact that $K$ depends on $\rho$, as well on the velocity field $\bfv$. 

\begin{figure}
\centering
\includegraphics[width=0.48\textwidth]{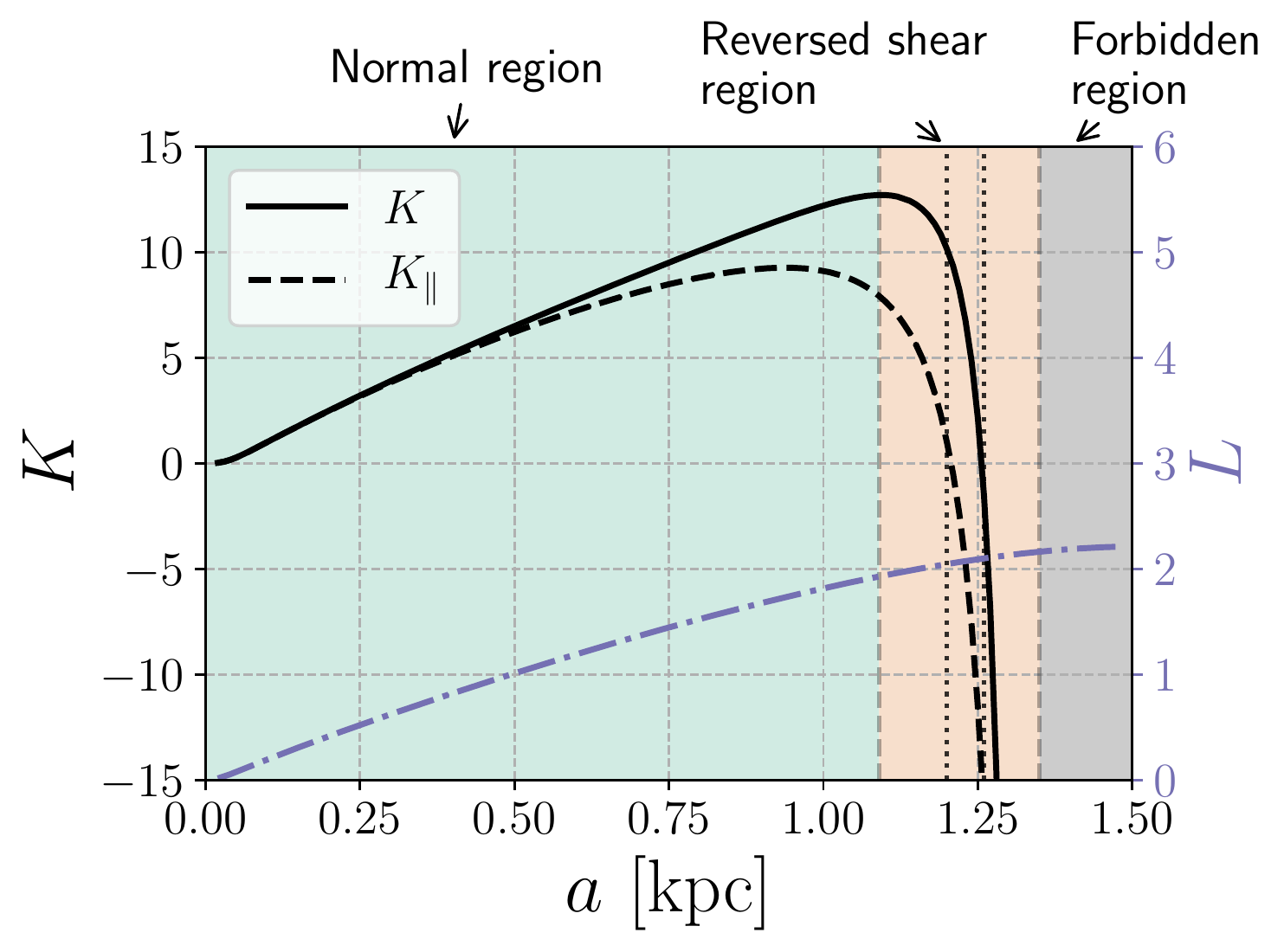}
\caption{\emph{Full black line:} the flux of angular momentum through orbit contours as a function of the semi-major axis (see Eq. \ref{eq:K}). \emph{Black dashed line:} the flux of angular momentum calculated taking into account only the force tangential to $x_2$ orbits and neglecting the component perpendicular to them (see Eq. \ref{eq:Kparallel}). Both $K$ curves are calculated under the assumption that $\nu=\rm constant$ and $\rho=\rm constant$ along the $y$ axis. Units are normalised such that $\rho=\nu=1$ and are $[K]=100 \kms \kpc $. \emph{Blue dashed line:} the angular momentum per unit mass of an infinitely thin annulus comprised between two $x_2$ orbits, defined as $L=\int  (\bfv \times \bfr) \di m/\int\di m$ where $\di m$ is a mass element along the annulus and $\bfv$ is the velocity in an inertial frame, as a function of its semi-major axis. Units are $[L] = 100 \kms \kpc$.}
\label{fig:SBM_3}
\end{figure}

The solid line in Fig. \ref{fig:SBM_3} plots the curve $K(a)$ calculated under the following assumptions:  
\begin{enumerate}
\item The velocity field is exactly that of $x_2$ orbits.
\item $\nu = \rm constant$.
\item The density distribution is of the type described by Eq. \eqref{eq:rhoorbit} with the mass scaling (i.e., the values of $\alpha$) chosen such that $\rho = \rm constant$ along the y axis. 
\end{enumerate}
From this figure we infer that if a disc is set up with this initial configuration, material in the region where $\di K/\di a>0$ will move inwards, while material where $\di K/ \di a<0$ will move outwards. Hence, we expect a gap to open approximately between the green and orange region, and a ring to form somewhere in the orange region. However, these considerations only apply at the initial instant. It is not easy to determine the subsequent development of the $K(a)$ curve in detail because i) the density distribution will change ii) viscous (and pressure, if present) forces will deform the orbits, so that the velocity field will deviate from that of $x_2$ orbits. Both these facts will affect the curve $K(a)$.

While a redistribution of $\rho$ changes the shape of the curve $K(a)$, the point $K=0$ seems of particular significance. It is clear from the definition \eqref{eq:K} that this point must remain fixed if the density is redistributed while being forced to remain a steady state of the type described by \eqref{eq:rhoorbit}. Under this assumption, also the regions where the curve $K(a)$ is positive (negative) do not change. Hence, we must always have $\di K /\di a<0$ at the point where $K=0$ and we expect an accumulation of mass just outside the $K=0$ orbit. Thus we might expect that a ring should form with approximately the size and shape of the orbit with semi-major axis $a_{\rm ring}$ such that $K(a_{\rm ring})=0$. 

However, the numerical experiments that we have performed show that this usually overestimates the size of the ring. The discrepancy is due to the fact that in the presence of viscous and pressure forces, even in the most controlled and idealised situations it is not exactly true that the density remains exactly of the type described by \eqref{eq:rhoorbit} and that the velocity remains exactly that of $x_2$ orbits. 

Numerical simulations suggest that a related quantity that seems to do a better job in predicting the size of the ring is 

\begin{equation} \label{eq:Kparallel}
\mathbf{K}_\parallel = \int \rho \mathbf{k}_\parallel \ \di l \;,
\end{equation}
where $\mathbf{k}_\parallel = \bfr \times \mathbf{f}_\parallel$ is the torque calculated using only the component of the viscous force that is parallel to the $x_2$ orbits and neglecting the component perpendicular to the orbit, i.e.
\begin{equation}
\mathbf{f}_\parallel = \left( \mathbf{f} \cdot \hatt \right) \hatt
\end{equation}
see Fig. \ref{fig:diagram} for the definition of $\hatt$. The dashed black line in Fig. \ref{fig:SBM_3} shows $K_\parallel$ for comparison with $K$. Numerical simulations suggest that a ring forms where $K_\parallel=0$. Note that in the circular case $\mathbf{k} = \mathbf{k}_\parallel$, since the force perpendicular to orbits is directed radially and produces no torque. Note also that in the non-circular case, $\mathbf{f}$ depends on the factor $2/3$ on the right side of \eqref{eq:sigmaij}, which is related to the ratio of shear to bulk viscosity, while $\mathbf{f}_\parallel$ is independent of this value.  

What is the justification for neglecting the component of the force perpendicular to the orbits? This component acts in the direction perpendicular to the orbits, so it works by pushing the contours of the orbits to deform them. This force does not try to `speed up' or `slow down' orbits, and does not correspond to our intuitive intuition of a viscous force which is related to layers of fluid sliding relative to each other. Thus it may be that this component acts, in essence, as some sort of pressure, and only causes a slight readjustment in the shape of the orbit until it becomes unimportant. This is speculative and ultimately based on the empirical fact that based on our numerical experiments the location $K_\parallel=0$ seems to be a good predictor in predicting where a ring forms (see Sect. \ref{sec:chemresults}). More tests with different potentials are needed to verify whether this is really true, and if so why.

\subsection{Viscous vs gravitational torques}

\begin{figure*}
\centering
\includegraphics[width=1.0\textwidth]{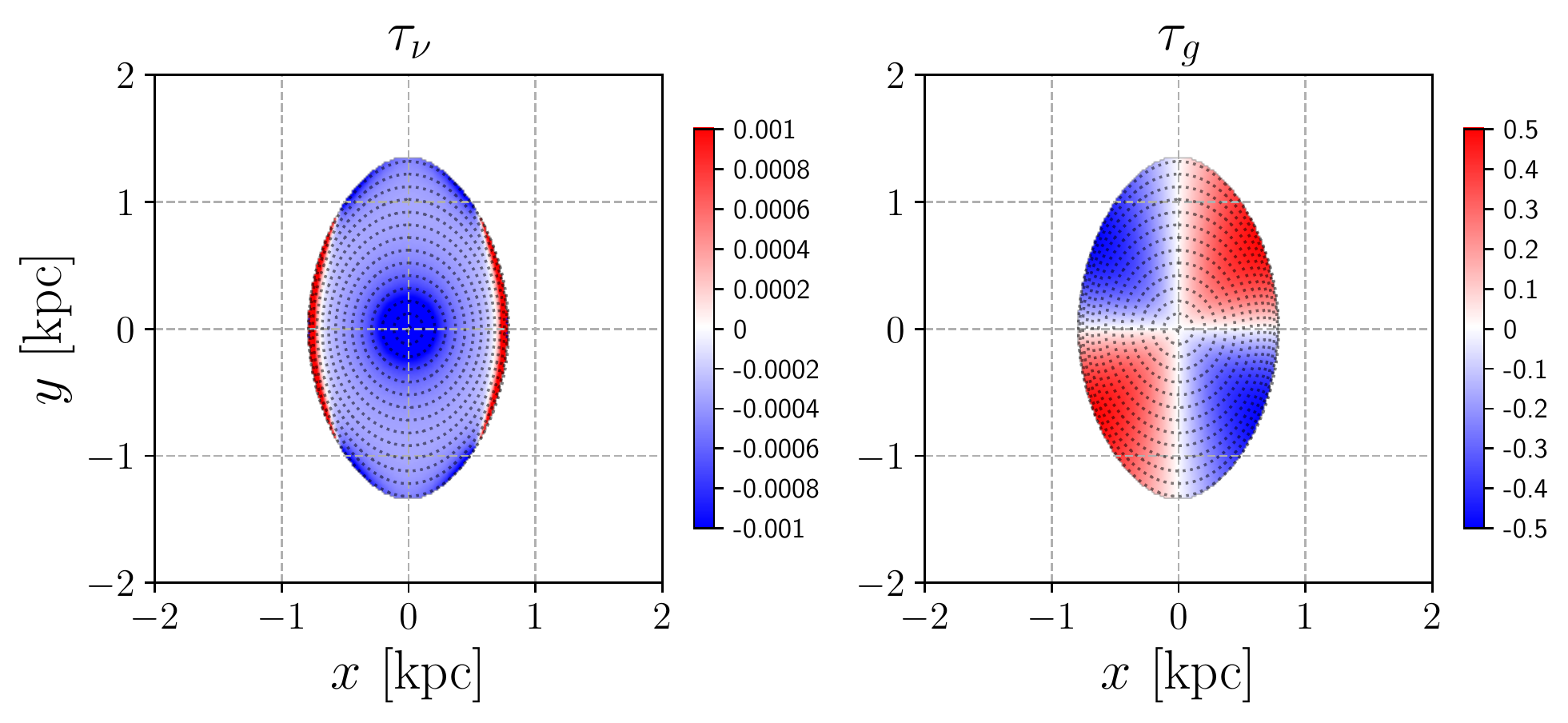}
\caption{Comparison of viscous and gravitational torques. \emph{Left:} viscous torques calculated using Eq. \eqref{eq:taunu} assuming that $\nu = 10^{-4} \times 100 \kms \kpc$ and that $\rho$ is of the type given by Eq. \eqref{eq:rhoorbit} with mass normalisation such that $\rho=\rm constant$ along the $y$ axis. \emph{Right:} gravitational torques per unit mass, see Eq. \ref{eq:taug}. Dots delineate the $x_2$ orbits (left panel of Fig. \ref{fig:SBM_1}). Units are $[\tau] = (100 \kms)^2$.}
\label{fig:torques}
\end{figure*}

It is instructive to compare gravitational and viscous torques. The right panel in Fig. \ref{fig:torques} shows what is sometimes referred to as the `butterfly diagram', i.e. it shows the gravitational torque per unit mass, given by:
\begin{equation} \label{eq:taug}
\tau_g = \frac{\pa \Phi}{\pa \theta}.
\end{equation}
The left panel shows the analogous quantity related to viscous forces, i.e. it shows the viscous torques per unit mass:
\begin{equation} \label{eq:taunu}
\tau_\nu = \frac{1}{\rho} \left\{ \pa_x \left[ \rho \left( x \sigma_{yx} - y \sigma_{xx} \right) \right] + \pa_y \left[ \rho \left( x \sigma_{yy} - y \sigma_{xy} \right) \right] \right\}.
\end{equation}
A derivation of Eq. \eqref{eq:taunu} is given in Appendix \ref{appendix:L}. The quantity $\tau_\nu$ differs from $k$ (see Eq. \ref{eq:k}) in that the latter is the torque exerted on contours, while the former is the net angular momentum acting on fluid elements and corresponds to $k$ integrated over an infinitesimally small closed contour. Unlike the gravitational torques, the viscous torques depend on the density and velocity distributions. The left panel in Fig. \ref{fig:torques} is constructed under the same assumptions used to construct the $K$ curves in Fig. \ref{fig:SBM_3}, namely $\rho= \rm constant$ along the $y$ axis and $\nu = \rm constant$. The numbers in the colourbar are calculated assuming a value of $\nu = 10^{-4} \times 100 \kms \kpc$, which is the same used for simulations in Section \ref{sec:resultsvisc}. This is a fairly low value which is a factor of $30$ below estimates of turbulent viscosity in the nuclear regions of real galaxies (see Sect. \ref{sec:visc}).

Blue regions in the left panel correspond to `normal shear', while the red sides in the left panel are regions of `reversed shear'. Gas orbiting in the blue region of the left panel looses angular momentum due to viscous torques, while gas in the red regions gains angular momentum. While the instantaneous value of gravitational torques is usually much larger than the instantaneous value of viscous torques, the time average of the former over an $x_2$ orbit is always zero, while the average of the latter is usually non-zero.

\section{Numerical experiments} \label{sec:experiments}

\subsection{Isothermal viscous rings} \label{sec:resultsvisc}

The first test that we perform is the analog of the standard viscous ring spreading problem (e.g. fig. 1 of \citealt{Pringle1981}) but instead of the Kepler potential we use our barred gravitational potential. The simulations are two-dimensional, isothermal and include an explicit viscosity term. The gas moves in a rigidly rotating externally imposed barred potential and self-gravity of the gas is neglected. Thus, the equations of motion in an inertial frame are the continuity equation and the Navier-Stokes equation
\begin{align}
\pa_t \rho + \pa_i \left(\rho v_i \right) & = 0, \label{eq:ns1} \\
\pa_t v_i + (v_j \pa_j ) v_i 		& =  - \frac{1}{\rho}\pa_i P  - \frac{1}{\rho} \pa_j (\rho \sigma_{ij}) - \pa_i \Phi, \label{eq:ns2}
\end{align}
where $\sigma_{ij}$ is given by Eq. \eqref{eq:sigmaij}, $\Phi$ is the external gravitational potential and $P = \cs^2 \rho$ where $\cs$ is a constant.

\subsubsection{Numerical code}

We use the public code {\sc Pluto} \citep{Pluto2007}. This is a free software for the numerical solution of systems of conservation laws targeting high Mach number flows in astrophysical fluid dynamics. We use a two-dimensional static polar grid in the region $R \times \theta = [0.1,2.0] \kpc \times [0, 2\pi]$. The grid is uniformly spaced in both $R$ and $\theta$ with $950 \times 1440$ points, which corresponds to a resolution of $2 \pc$ in radius and $0.25 \degree$ in polar angle. We use the following parameters: {\sc RK2} time-stepping, no dimensional splitting, {\sc Hll} Riemann solver and the {\sc Van Albada} flux limiter. Boundary conditions are reflective on the inner boundary at $R=0.1\kpc$ and outflow on the outer boundary at $R=2.0\kpc$. Viscosity is treated with the {\sc Explicit} scheme.

\subsubsection{Initial conditions}

The initial density distribution is taken to be of the type described by Eq. \eqref{eq:rhoorbit}. The function $\alpha$ is chosen so that the density distribution along the $y$ axis is the sum of one or more gaussians of the following type:
\begin{equation}
\rho_0(a) = \exp\left[ - \left(\frac{a - a_0}{\Delta a}\right)^2 \right]
\end{equation}
where $\Delta a = 50 \pc$ and $a_0$ is a parameter that controls where the ring is centred. The initial velocity distribution is taken to be that of $x_2$ orbits in runs with non-axisymmetric potentials, or the circular velocity $v_{\rm circ}(R) = \sqrt{R \di \Phi_0/\di R}$ in runs with axisymmetric potentials. In regions where $x_2$ orbits do not exist, we have used a nearest-neighbour interpolation to assign the velocity of the closest $x_2$ orbit. This makes no difference in practice because those regions were always initially void of gas, and the density was set to a negligibly small value of $\rho = 10^{-12}$. The density units are arbitrary since the equations of motions \eqref{eq:ns1} and \eqref{eq:ns2} are invariant under density rescaling. 

\subsubsection{Results}

\begin{figure*}
\centering
\includegraphics[width=1.0\textwidth]{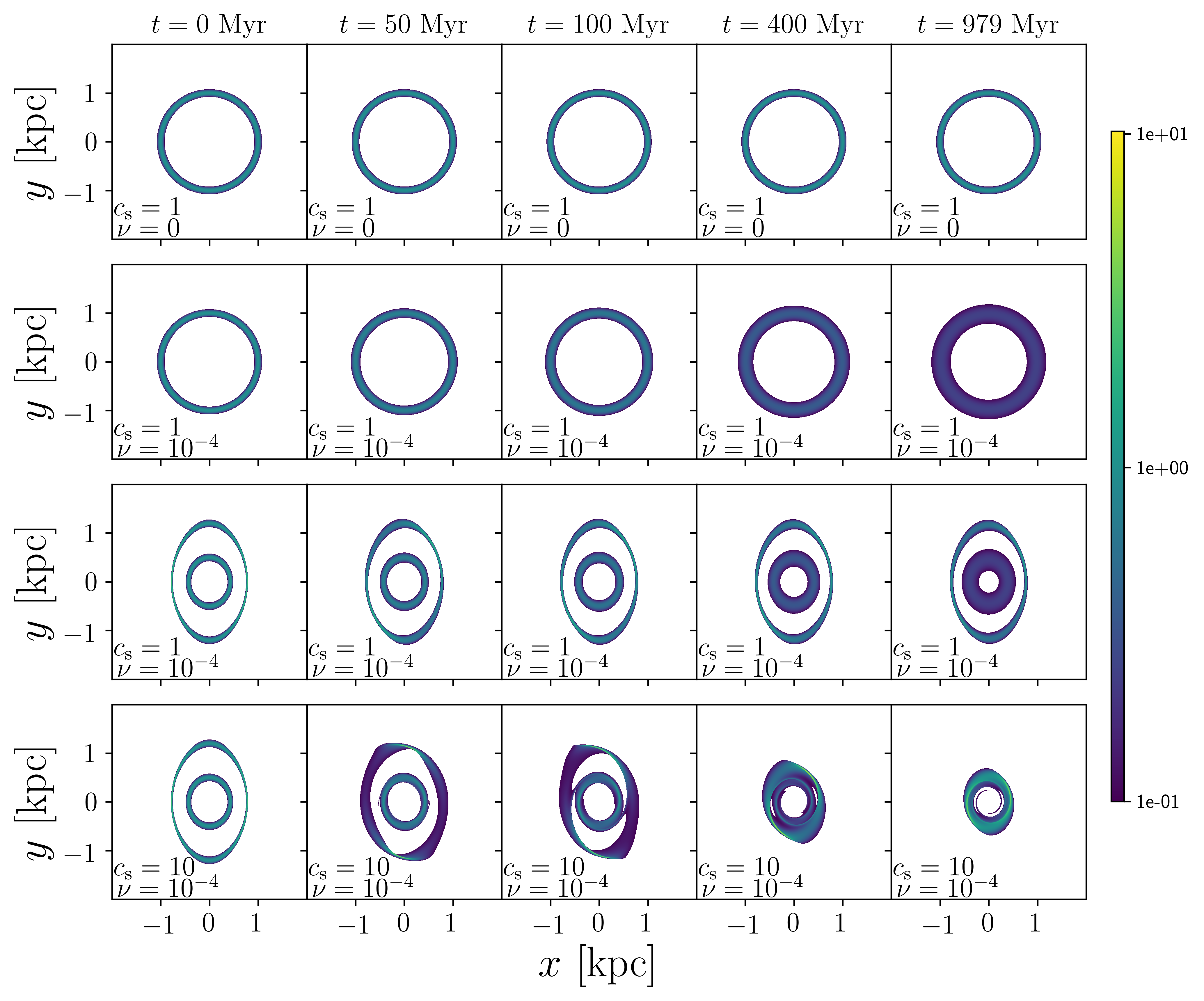}
\caption{Simulations of gaseous rings for different potentials and values of $\cs$ and $\nu$. Each panel shows the gas surface density and time increases from left to right as indicated at the top of the figure. \emph{First row:} one ring is placed in a purely axisymmetric potential (Eq. \ref{eq:log}) and viscosity is turned off, $\nu=0$. The ring does not spread. \emph{Second row:} same as the first row, but viscosity is set to a value $\nu = 10^{-4} \times 100 \kms \kpc  \simeq 3 \times 10^{24} {\rm cm^2\ s^{-1}}$. The ring spreads due to viscous torques. \emph{Third row:} two rings are placed in a barred potential given by \eqref{eq:sbmphi} in the presence of viscosity. The inner ring (placed in the green region in Fig. \ref{fig:SBM_1}) spreads, similarly to the axisymmetric case in the second row, while the outer ring (placed in the orange region in Fig. \ref{fig:SBM_1}) does not spread despite the presence of viscosity. \emph{Fourth row:} same as third row, but for a higher value of the sound speed $\cs$. The ring is destroyed by spirals created by pressure. In all simulations the initial conditions are given such that in the absence of pressure and viscous forces it would be a steady state. Gas circulates clockwise and the bar major axis is horizontal.}
\label{fig:rho1}
\end{figure*}

\begin{figure}
\centering
\includegraphics[width=0.48\textwidth]{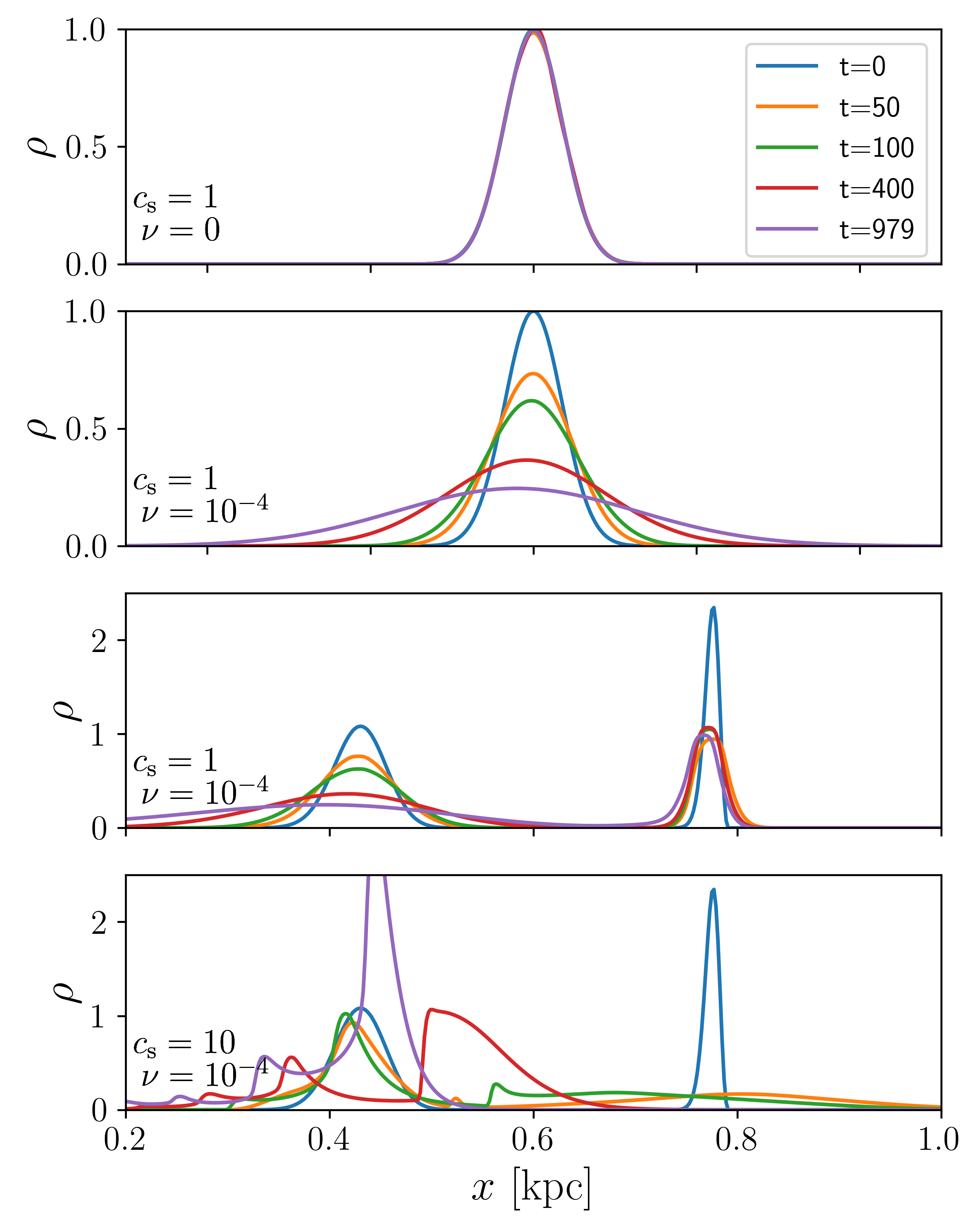}
\caption{Density cuts along the $x$ axis for the same simulations shown in Fig. \ref{fig:rho1}.}
\label{fig:rho3}
\end{figure}

Fig. \ref{fig:rho1} shows the time evolution of the 2D surface density for four different simulations with different potentials (barred or axisymmetric) and different values of $\nu$ and $\cs$. Fig. \ref{fig:rho3} shows cuts of the surface density along the $x$ axis for the same snapshots.

The first row shows the evolution of a ring in an axisymmetric potential when $\nu=0$. The ring is centred on $a_0=1\kpc$ and the potential is given by Eq. \eqref{eq:log} (see the circular velocity curve in Fig. \ref{fig:SBM_2}). The sound speed is $\cs = 1\kms$. As expected, in the absence of viscosity the ring does not spread. Not much happens. This also shows that the numerical viscosity is negligibly small.

The second row shows what happens when we turn on the viscosity in the axisymmetric case. Everything is the same as in the first row except that viscosity now has a value $\nu = 10^{-4} \times 100 \kms \kpc  \simeq 3 \times 10^{24} {\rm cm^2\ s^{-1}}$. Viscosity makes the ring spread, exactly as we would expect in analogy to the standard viscous ring spreading problem (compare for example the second row in Fig. \ref{fig:rho3} with fig. 1 of \citealt{Pringle1981}). The value of viscosity used here is fairly low, roughly a factor of $30$ lower than estimates for real galaxies based on turbulent viscosity (see Sect. \ref{sec:visc}). As we will discuss in Sect. \ref{sec:visc}, this shows that a relatively low value of $\nu$ can already produce significant effects over a time scale of $1 \Gyr$.

The third row shows the evolution of two rings in the same barred potential used to construct Fig. \ref{fig:SBM_1}. The axisymmetric part $\Phi_0$ of this potential is exactly the same as in the first two rows, but in addition now also the quadrupole $\Phi_2$ is present (see the blue dashed line in Fig. \ref{fig:SBM_2} and Eq. \ref{eq:sbmphi}). The two rings are initially centred on $a_0=0.4 \kpc$ and $a_0=1.2\kpc$ respectively, i.e. the inner one is in the middle of the green region in Fig. \ref{fig:SBM_1} while the outer one is in the middle of the orange region. The inner ring spreads, similarly to the circular case (second row). The outer ring instead, after an initial adjustment, does not spread and at $t= 979 \Myr$ still has the same width it had at $t= 50 \Myr$ (see third row in Fig. \ref{fig:rho3}). This is in remarkable agreement with the idea described in Sect. \ref{sec:theory}.

The fourth row shows a simulation which is identical to that of the third row except that the sound speed is increased to $\cs=10\kms$. This time, the outer ring does not survive. Pressure forces quickly create spiral shocks, which destroy the outer ring and make material on the outer ring plunge inwards and merge with the inner ring. Thus whether the mechanism discussed in Sect. \ref{sec:theory} can work depends on the relative importance of pressure forces. In order to understand the effects of pressure we have run a further isothermal simulation which is described in the next section. The effects of pressure will be discussed in Sect. \ref{sec:pressure}.

\subsection{Isothermal disc} \label{sec:results2}

In order to better understand the effects of pressure, we have run a further isothermal simulation in which we have set up a filled disc of gas on $x_2$ orbits. The numerical setup is identical to that of Sect. \ref{sec:resultsvisc}, except for the initial conditions. The sound speed of this simulation is $\cs=10 \kms$ while the viscosity is turned off, $\nu=0$.

\subsubsection{Initial conditions}

The initial density distribution is such that in the absence of pressure and viscous forces it would be a steady state of the type described by Eq. \eqref{eq:rhoorbit}. The normalisation $\alpha$ is chosen so that the density distribution along the $y$ axis is:
\begin{equation}
\rho_0(a) = \begin{cases}
 1 & \text{if} \qquad a \leq 1.35 \kpc , \\
 0 &  \text{if} \qquad a>1.35 \kpc . \\
\end{cases}
\end{equation}
The value $a=1.35\kpc$ corresponds to the orbit at which the forbidden region starts (cf. the grey area in Fig. \ref{fig:SBM_1}).

\subsubsection{Results}

\begin{figure*}
\centering
\includegraphics[width=1.0\textwidth]{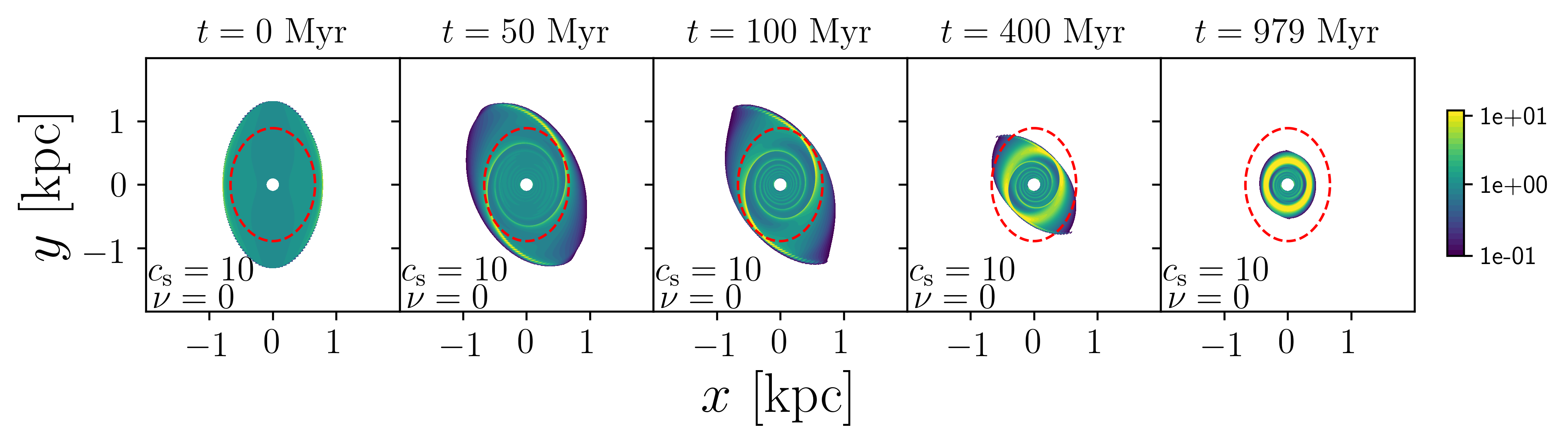}
\caption{Evolution of the surface density for an idealised isothermal simulation with $\cs=10\kms$, $\nu=0$. Initial conditions are given such that $\rho=1$ along the $y$ axis and that in the absence of pressure and viscous forces it would be a steady state. The dashed red line shows the orbit examined in more detail in Figure~\ref{fig:tube}.}
\label{fig:rho5}
\end{figure*}

\begin{figure}
\centering
\includegraphics[width=0.48\textwidth]{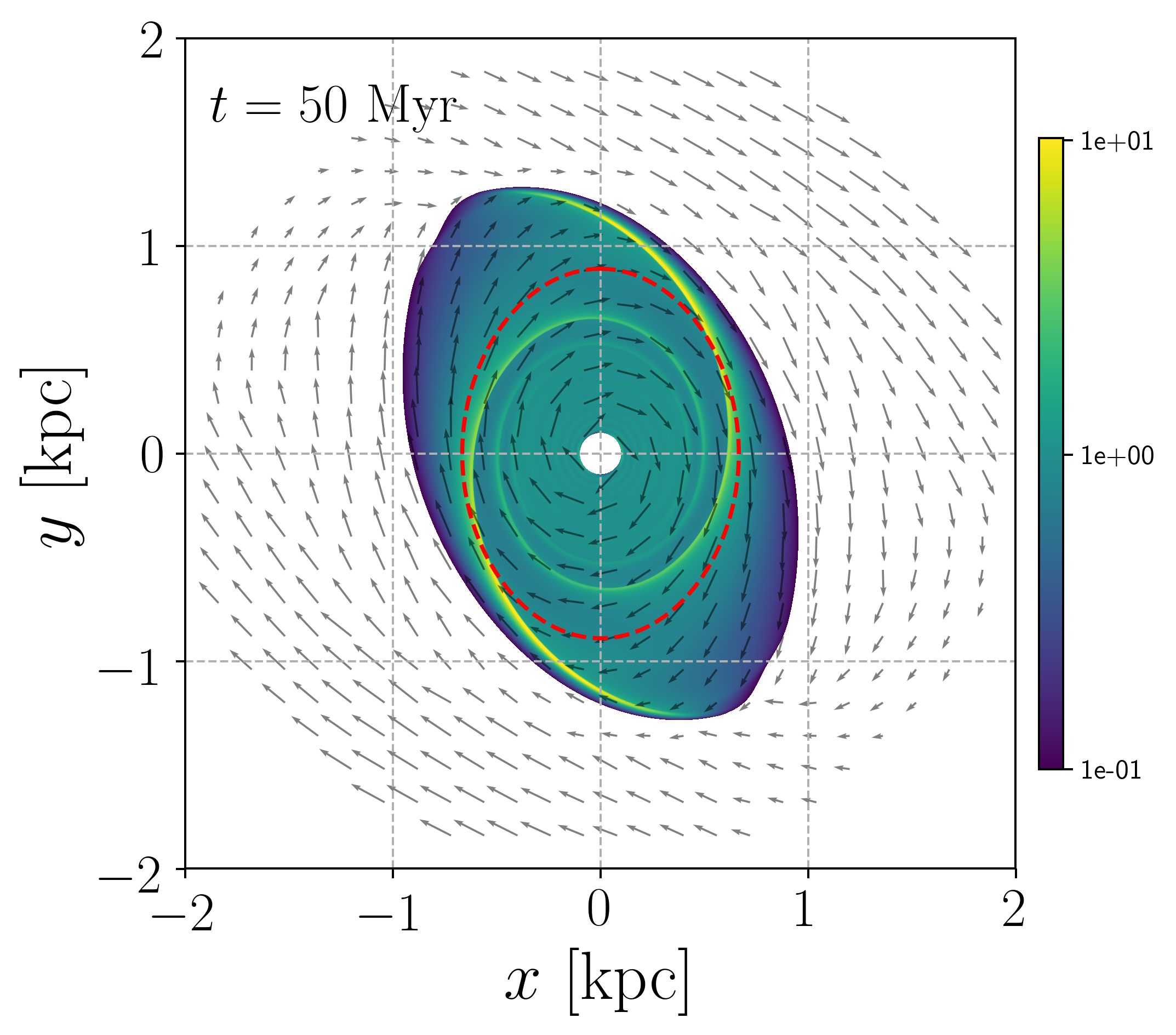}
\caption{Velocity field superimposed on the surface density for the snapshot at $t=50 \Myr$ of the simulation shown in Fig. \ref{fig:rho5}.}
\label{fig:rho6}
\end{figure}

\begin{figure}
\centering
\includegraphics[width=0.48\textwidth]{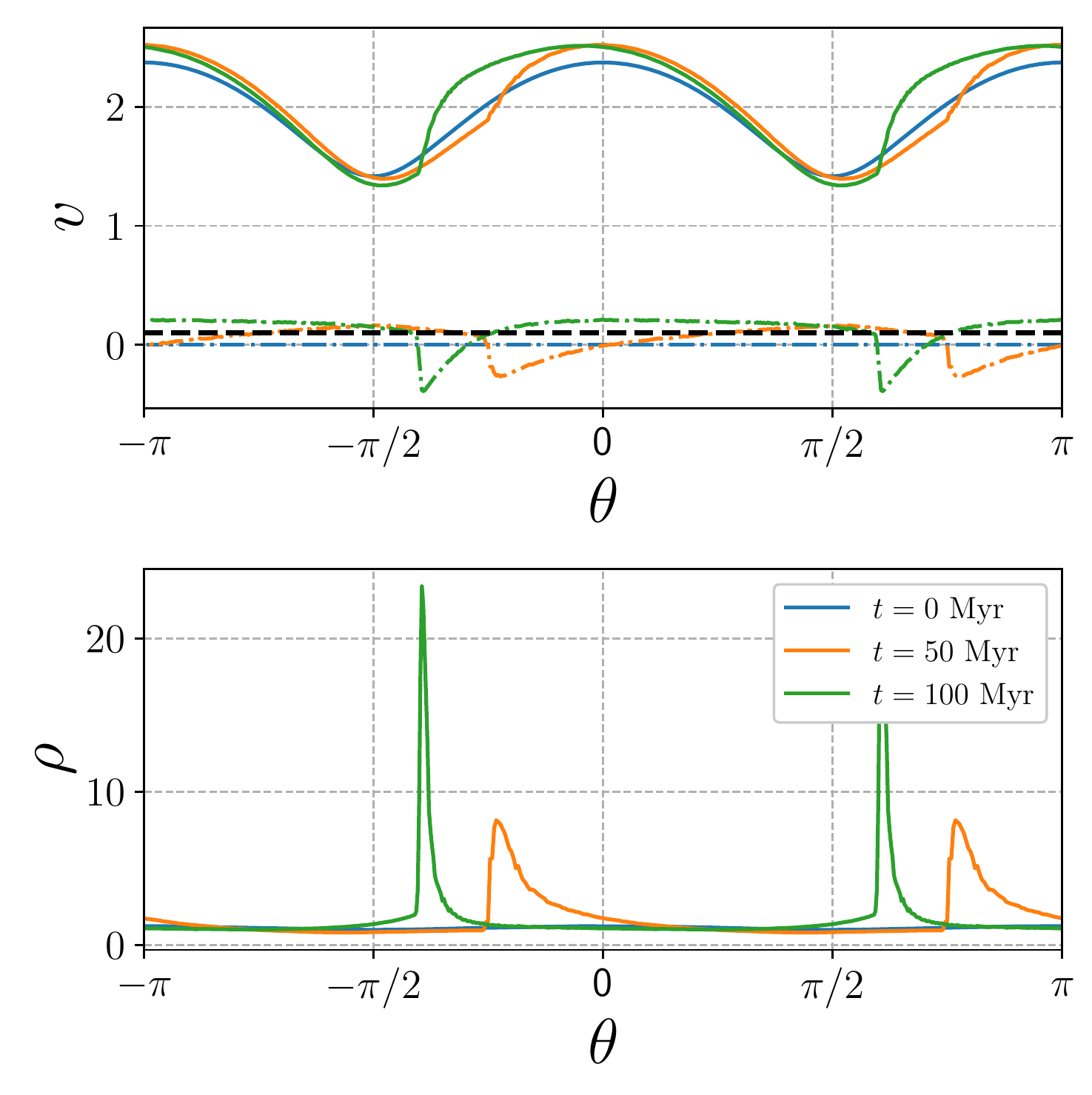}
\caption{Velocity and density along the dashed $x_2$ orbit in Fig. \ref{fig:rho5} and \ref{fig:rho6}. In the top panel, the full lines are the component of the velocity parallel to the orbit, the dot-dashed lines are the components of the velocity perpendicular to the orbit and the black dashed line indicates the value of the sound speed.}
\label{fig:tube}
\end{figure}

Fig. \ref{fig:rho5} shows the time evolution of the surface density, while Fig. \ref{fig:rho6} shows in more detail the snapshot at $t=50 \Myr$. Pressure forces cause the shape of the gas streamlines to deviate from $x_2$ orbits almost immediately, creating spiral shocks shortly after the start of the simulation. That these are really shocks as opposed to generic density waves can be seen from Fig. \ref{fig:tube}, which shows cuts of the density and the velocity along the red dashed orbit shown in Figs. \ref{fig:rho5} and \ref{fig:rho6}. These cuts show that density and velocity have discontinuous jumps, which is the defining characteristics of shocks. They also show that the shock front moves upstream with respect to the overall flow (i.e., it moves to the left in Fig. \ref{fig:tube}). At the resolution of the simulation shown in Fig. \ref{fig:rho5} the spiral shocks appear smooth, but we have verified that at higher resolution the shock fronts develop wiggles and break up as a consequence of the wiggle instability \citep{WadaKoda2004,KimKimKim2014,Sormani+2017}. This instability is not important for the discussion in this paper. 

A notable feature of the simulation that is clearly visible in Fig. \ref{fig:rho5} is that the disc shrinks with time. This happens because for an extended period of time, which lasts approximately for $0 \lesssim t  \lesssim  800 \Myr$, the overall surface density distribution is not symmetric with respect to the $y$ axis but is tilted so that so that much more material lies in the two quadrants $(x<0,y>0)$ and $(x>0,y<0)$ than in the other two quadrants (for example, it is clear that at $t= 50 \Myr$ the major axis of the gas disc makes an angle with the $y$ axis and points towards negative $x$). Since gravitational torques from the bar potential act in such a way as to remove angular momentum from material in these two quadrants, the total angular momentum steadily decreases during the tilted phase and the disc shrinks. After this phase the disc returns to a more symmetric configuration (see the final panel at $t=979 \Myr$ in Fig \ref{fig:rho5}) and the angular momentum stabilises to a value approximately $60\%$ of the initial value. Now the bar is unable to remover further angular momentum, and the final state has a fat ring at with $R\sim 0.4\kpc$, but this ring has nothing to do with the mechanism described in Sect. \ref{sec:theory}.\footnote{The final state will slowly shrink over time due to viscous torques acting as in a standard accretion disc, but this shrinking is extremely slow because numerical viscosity is extremely small (first row of Fig. \ref{fig:rho1}).} The physical reasons that cause the tilt and hence the removal of angular momentum can be traced back to the spiral shocks, see Sect. \ref{sec:pressure}.

Note that based on the resonant theory, we might have naively expected exactly the opposite, i.e. that the gas \emph{gains} angular momentum, because the spiral pattern is mostly in the quadrants $(x>0,y>0)$ and $(x<0,y<0)$, where the torque is positive. This shows that expectations based on the naive interpretation of the torques acting on the spiral pattern should be taken with great caution, and explains why predictions of the resonant theory are often not verified in actual simulations \citep{Kim++2012a}.

We have repeated this simulation using a value of the viscosity $\nu = 10^{-4} \times 100 \kms \kpc$, but the results are almost indistinguishable. The effects of pressure in this case completely overwhelm the effects of viscosity, which is negligible. This also shows that what creates the spirals is pressure, not viscosity. We continue the discussion of the effects of pressure in Sect. \ref{sec:pressure}.

\subsection{Non-isothermal simulation} \label{sec:chemresults}

The simulations discussed in the previous two sections are idealised numerical experiments and are far from realistic. As a more realistic example let us consider here the simulations of \cite{Sormani2018}. These authors used high resolution, three-dimensional hydrodynamical simulations performed using the moving-mesh code {\sc Arepo} \citep{Springel2010} to simulate the central regions of the Milky Way. In these simulations gas moves in a multi-component externally imposed rotating barred gravitational potential which is tuned to reproduce the characteristics of our Galaxy. The simulation box includes the whole region $R\leq 10 \kpc$ and thus the entire bar region, not just the nuclear region.

A key feature of the simulations of \cite{Sormani2018} is that they include a time-dependent chemical network that keeps track of the chemical composition of the gas (essentially hydrogen, carbon and oxygen chemistry). The chemical network is deeply coupled to the heating \& cooling processes of the interstellar medium (ISM). Thus, instead of assuming a simple equation of state, the temperature of the gas is calculated using a cooling function that takes into account several sources of heating \& cooling, which include contributions that depend on the chemical composition of the gas and the instantaneous rate of chemical reactions (e.g. H collisional excitation and ionisation, H$^{+}$ recombination, H$_2$ formation, etc), and external contributions (e.g. heating from the average UV interstellar radiation field, cosmic rays).\footnote{See Tables 1 and 2 in \cite{Sormani2018} for the full list of reactions included in the chemical model and processes included in the cooling function.} The net result is a two-phase ISM made of a cold component with $T\simeq 100 \rm K$ (which corresponds to $\cs \lesssim 1\kms$) and a warm component with $T\simeq10^4 \rm K$ (which corresponds to $\cs \simeq 10 \kms$).

The large-scale dynamics in these simulations follows the typical dynamics of gas flowing in a barred potential \citep[e.g.][]{Sellwood1993}. They are set up as follows.\footnote{Movies are available at \url{http://www.ita.uni-heidelberg.de/~mattia/videos/asymmetry/high/rhoproj_barframe.mp4} (whole bar region) and \url{http://www.ita.uni-heidelberg.de/~mattia/videos/asymmetry/high/rhoprojzoom_barframe.mp4} (zoom on the nuclear region).} To avoid transients, the gas is initially on circular orbits and the bar is introduced gradually during the first $150 \Myr$. As the bar grows, material in the outer parts ($R\gtrsim 3\kpc$) deviates from circular orbits and starts following $x_1$ orbits, which is a family of orbits elongated parallel to the bar. Two narrow large-scale shocks are formed along which gas plunges in a dynamical time from $R\sim3\kpc$ down to the nuclear region, where it settles on $x_2$ orbits.

\begin{figure}
\centering
\includegraphics[width=0.5\textwidth]{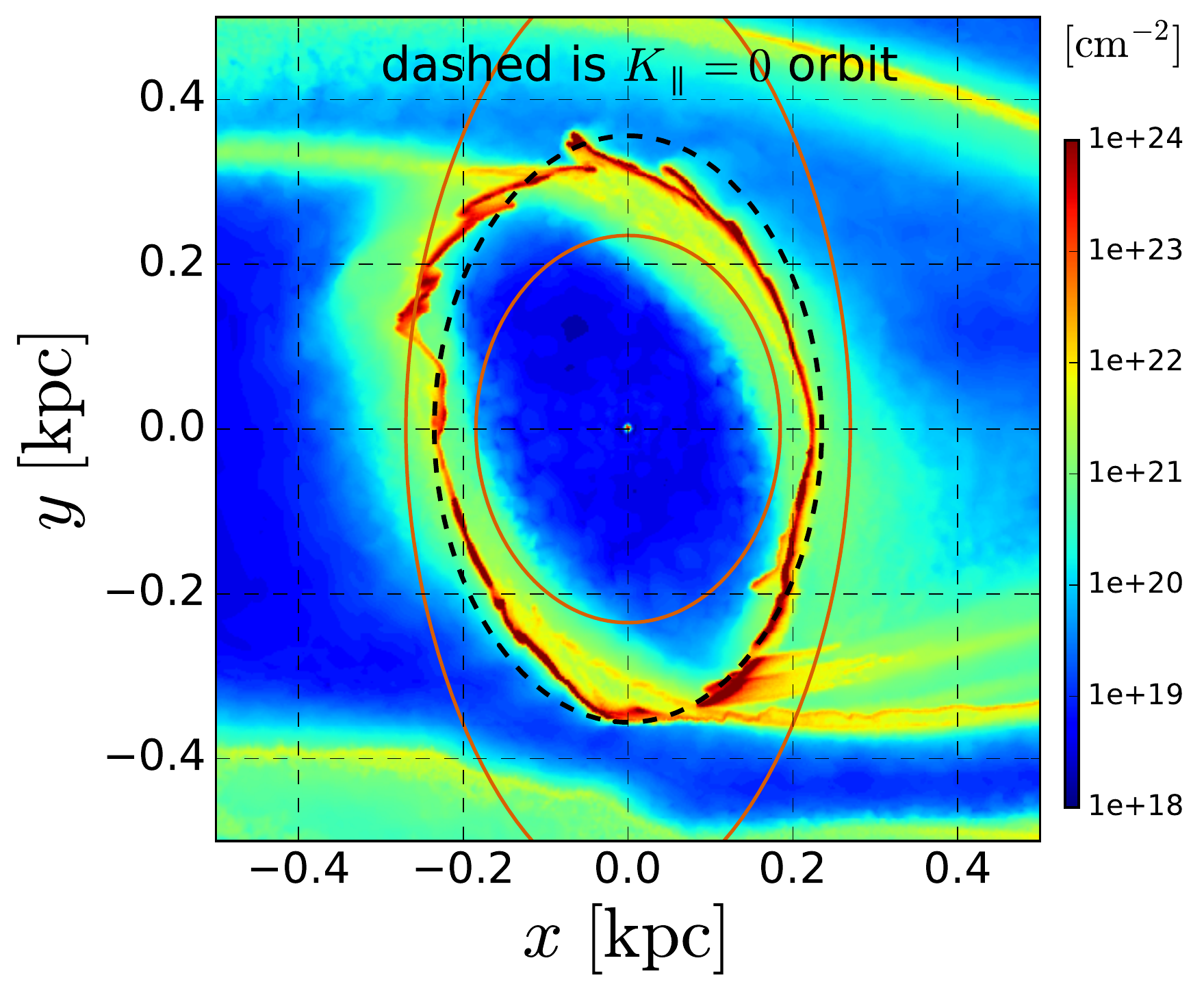}
\caption{Total surface density at $t=294 \Myr$ of the `high' simulation of \protect\cite{Sormani2018}. The dashed line corresponds to the $K_\parallel=0$ orbit. The two orange orbits are the orbits that delimit the orange region in Fig. \ref{fig:ridley_1}. Gas rotates clockwise.}
\label{fig:chem}
\end{figure}

A nuclear ring forms in these simulations with semi-major axis $\sim 400$ pc. This is shown in Fig. \ref{fig:chem} (see also last panel in fig. 7 of \citealt{Sormani2018}). It has the following striking characteristics:
\begin{enumerate}
\item The ring is extremely dense and thin.
\item The ring does not spread over time.
\item The ring is extremely long lived and its radius does not change in time even if the simulation is continued to $t \simeq 1 \Gyr$.
\end{enumerate}
Thus the ring seems to have all the characteristics expected from the mechanism described in Sect. \ref{sec:theory}. Indeed this is not a coincidence, since the study in this paper was inspired by the simulations of \cite{Sormani2018}. Note also that in the simulations of \cite{Sormani2018} self-gravity of the gas is ignored, so the ring self-gravity cannot play a role in determining its thinness.  

\begin{figure*}
\centering
\includegraphics[width=1.0\textwidth]{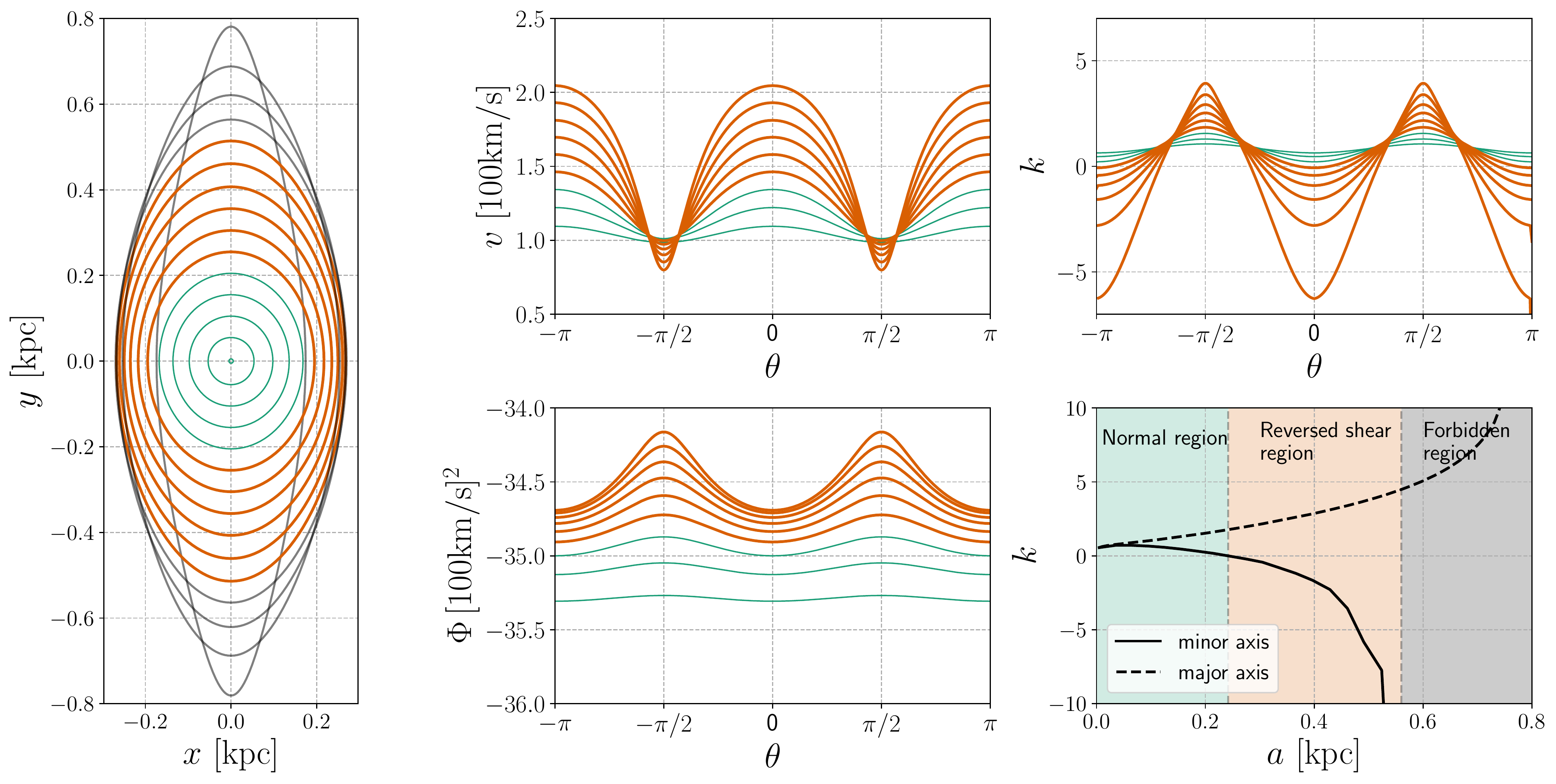}
\caption{Same as Fig. \ref{fig:SBM_1} but for the gravitational potential of \protect\cite{Sormani2018}.}
\label{fig:ridley_1}
\end{figure*}

\begin{figure}
\centering
\includegraphics[width=0.48\textwidth]{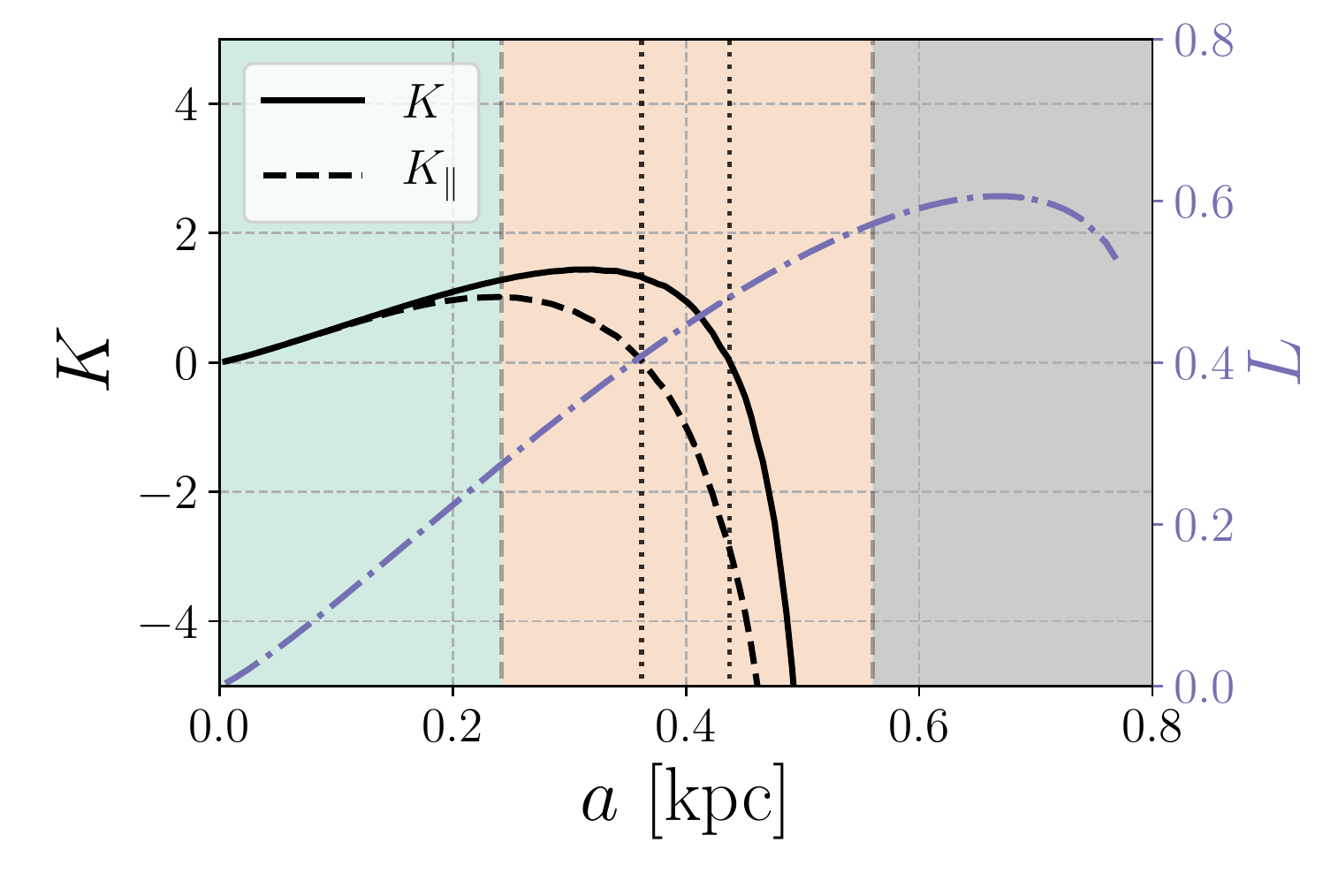}
\caption{Same as Fig. \ref{fig:SBM_3} but for the gravitational potential of \protect\cite{Sormani2018}.}
\label{fig:ridley_3}
\end{figure}

\begin{figure*}
\centering
\includegraphics[width=1.0\textwidth]{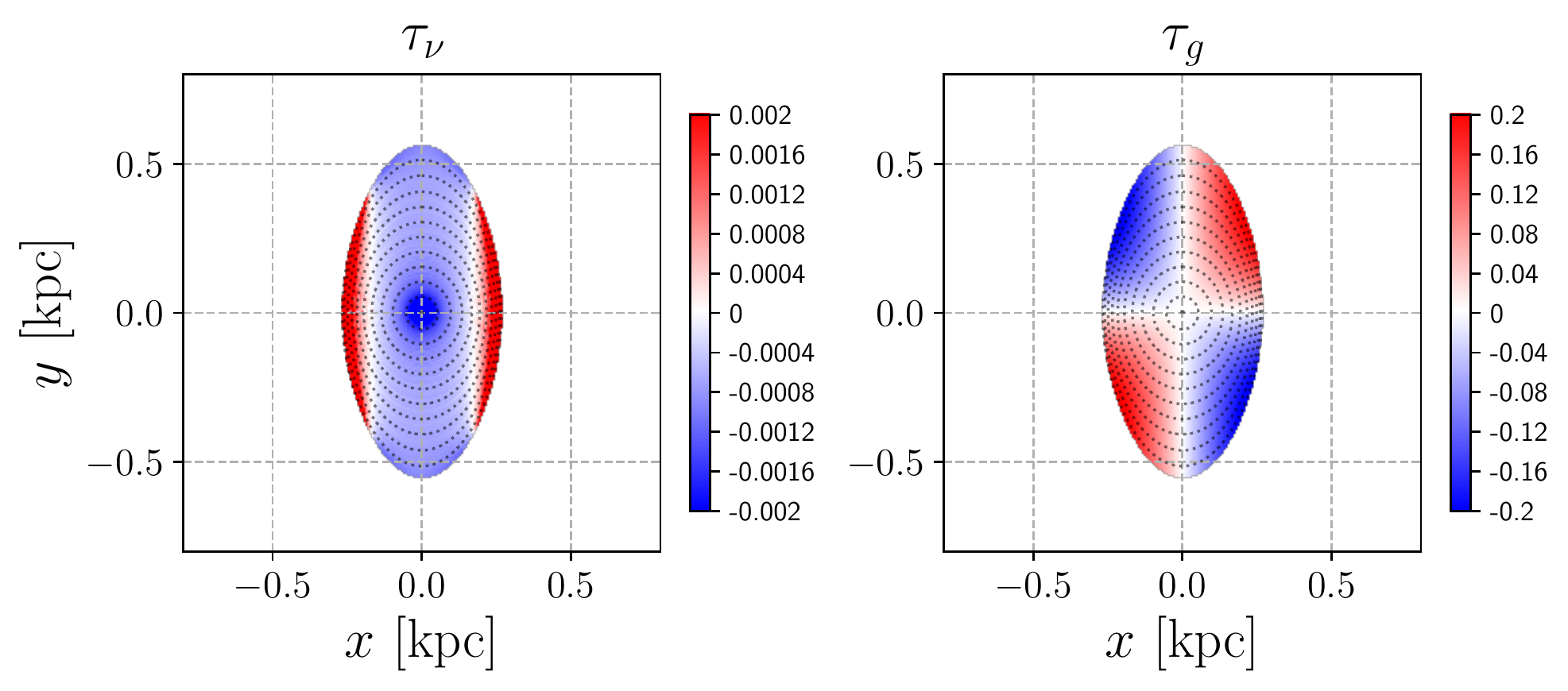}
\caption{Same as Fig. \ref{fig:torques} but for the gravitational potential of \protect\cite{Sormani2018}.}
\label{fig:torqueridley}
\end{figure*}

Figs. \ref{fig:ridley_1}, \ref{fig:ridley_3} and \ref{fig:torqueridley} show the $x_2$ orbits, the curve $K$ and the torques for the potential used in \cite{Sormani2018}. In accordance with the discussion in Sect. \ref{sec:theory}, the ring forms in the middle of the reversed shear region (orange in Fig. \ref{fig:ridley_1} and delimited by two orange orbits in Fig. \ref{fig:chem}). Fig. \ref{fig:chem} shows that the $K_\parallel=0$ orbit fits very well with both the size and the shape of the ring. This is a promising confirmation of the considerations in Sect. \ref{sec:predictions}, but more tests with different potentials are needed to be certain that this is not a coincidence.

\section{Discussion} \label{sec:discussion}

\subsection{Viscosity}
\label{sec:visc}

We have assumed that viscosity is present in our theory, but so far we have said nothing about the source of this viscosity. Let us first give a rough estimate of the minimum amount of viscosity needed to produce significant effects. Viscous stresses can significantly affect the dynamics of the system if the viscous time scale, $t_\nu \approx R^2/\nu$ (where $R$ is the typical dimension of the ring), is shorter than the Hubble time, $t_{\rm H}=14\text{ Gyr}$. This gives a minimum coefficient of
\begin{equation} \label{eq:numin}
\nu_{\rm min}\approx\frac{R^2}{T_{\rm H}}= 2\times 10^{25}\left(\frac{R}{1\text{ kpc}}\right)^2 \left(\frac{t_{\rm H}}{14\text{ Gyr}}\right)^{-1}\text{ cm$^2$ s$^{-1}$}.
\end{equation}

What is the viscosity of the real ISM? The molecular viscosity is negligible. But there are random turbulent motions of interstellar clouds, maintained by stellar feedback and MHD instabilities \citep[e.g.][]{SellwoodBalbus1999,MacLowKlessen2004}. These give rise to turbulent viscosity. Using the following typical parameters, namely (i) a cloud size $a\sim 10\text{ pc}$; (ii) a velocity dispersion $c\sim 8\text{ km s$^{-1}$}$; (iii) a fraction $f\sim 0.1$ of the volume filled by clouds, \cite{LyndenBellPringle1974} estimated the turbulent viscosity coefficient as
\begin{equation} \label{eq:nuism}
\nu\approx\frac{1}{3}c\frac{a}{f}=8\times 10^{25}\text{ cm$^2$ s$^{-1}$}.
\end{equation}
Since $\nu>\nu_{\rm min}$, we conclude that viscous stresses play an important role in shaping the gas distribution in the central $\sim 1\text{ kpc}$ of the Galaxy.

Note that the value given by \eqref{eq:numin} is roughly a factor of $10$ greater than the value $\nu  \simeq 3 \times 10^{24} {\rm cm^2\ s^{-1}}$ assumed for the simulations in Sect. \ref{sec:resultsvisc}, which was already sufficient to cause a significant spreading of an axisymmetric ring over a timescale of $~1 \Gyr$ (see Fig. \ref{fig:rho1}). This suggests viscosity may in general be more effective than naive estimates such as the one above indicate. This fact is also implicitly contained in the definition of the dimensionless parameter $\tau = 12 \nu t R_0^{-2} $ of \cite{Pringle1981} (see his equation 2.13), which plays the role of time in the analytical solution of the viscous spreading problem.\footnote{The solution given by \cite{Pringle1981} is the Green function of the problem.} The numerical factor $12$ means that viscosity acts $12$ times faster in that problem than the estimate $t_\nu \sim  R_0^2 / \nu$ would suggest.

Two more facts corroborate the conclusion that viscosity must be dynamically significant in the nuclear regions of galaxies. First, as already noted by \citeauthor{LyndenBellPringle1974}, the turbulent viscosity in galactic centres is likely to be even higher than their estimate due to the higher velocity dispersion of molecular clouds there \citep[e.g.][]{HeyerDame2015}. Second, the above discussion neglects another source of dissipation, namely shocks. At the locations of `reversed shear' along $x_2$ orbits, gas streamlines converge, creating a zone of enhanced interaction where material is compressed and shocked. Shocks are a natural way of producing dissipation, and it may be that this contributes in driving the gas towards the local minimum of the energy represented by the ring configuration.

\subsection{Understanding the effects of pressure} \label{sec:pressure}

The results of Sect. \ref{sec:resultsvisc} and \ref{sec:results2} suggest that at low sound speed ($\cs=1\kms$) a ring created by the mechanism described in Sect. \ref{sec:theory} can survive, while at high sound speed ($\cs = 10\kms$) the ring cannot survive because it is destroyed by spiral shocks. What is the origin of these spiral shocks? 

To understand this, it is useful to consider a very simplified toy problem. Gas on an $x_2$ orbit revolves periodically around the galactic centre and feels a periodic potential that is an oscillating function of $\theta$ (see the middle-bottom panel of Fig. \ref{fig:SBM_1}). The gas is faster (slower) where the potential is low (high) (see the middle-top panel of Fig. \ref{fig:SBM_1}). Very crudely, we may model this gas as a one-dimensional isothermal fluid that moves in a periodic potential given by: 
\begin{align}
		\Phi(x) & =  - 2 \cos(2 \pi x).
\end{align}
The equations of motion are then the one-dimensional continuity and Euler equations:
\begin{align}
\pa_t \rho + \pa_x \left( \rho v \right)				& = 0 \\
\pa_t v + v \, \pa_x v							& = - \frac{\pa_x P}{\rho} - \pa_x \Phi
\end{align}
where $P = \cs^2 \rho$. In the absence of the pressure term ($\cs=0$), it is easy to check that the following is a steady state of the system:
\begin{align}
v(x) & = \sqrt{2(B - \Phi(x))} \label{eq:toysteady1} \\
\rho(x) & = 1/v(x) \label{eq:toysteady2} 
\end{align}
where $B$ is a constant. This corresponds to a purely ballistic state, i.e. it is the toy problem version of gas on an $x_2$ orbit.

\begin{figure}
\centering
\includegraphics[width=0.48\textwidth]{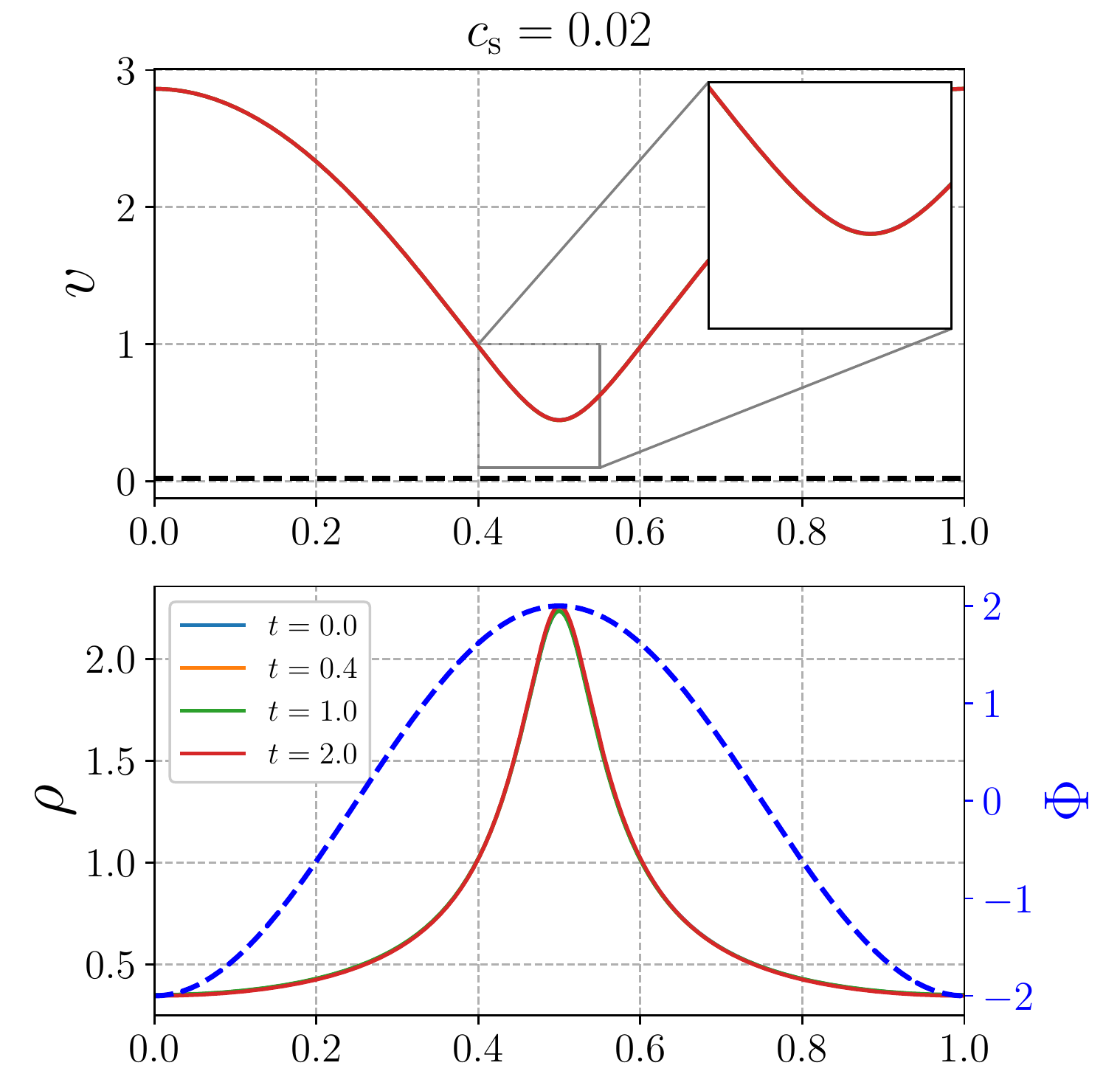}
\caption{Evolution of density and velocity in the 1D toy problem described in Section \ref{sec:pressure} for a low value of the sound speed, $\cs=0.02$. No shocks are formed.}
\label{fig:tube_cs002}
\end{figure}

\begin{figure}
\centering
\includegraphics[width=0.48\textwidth]{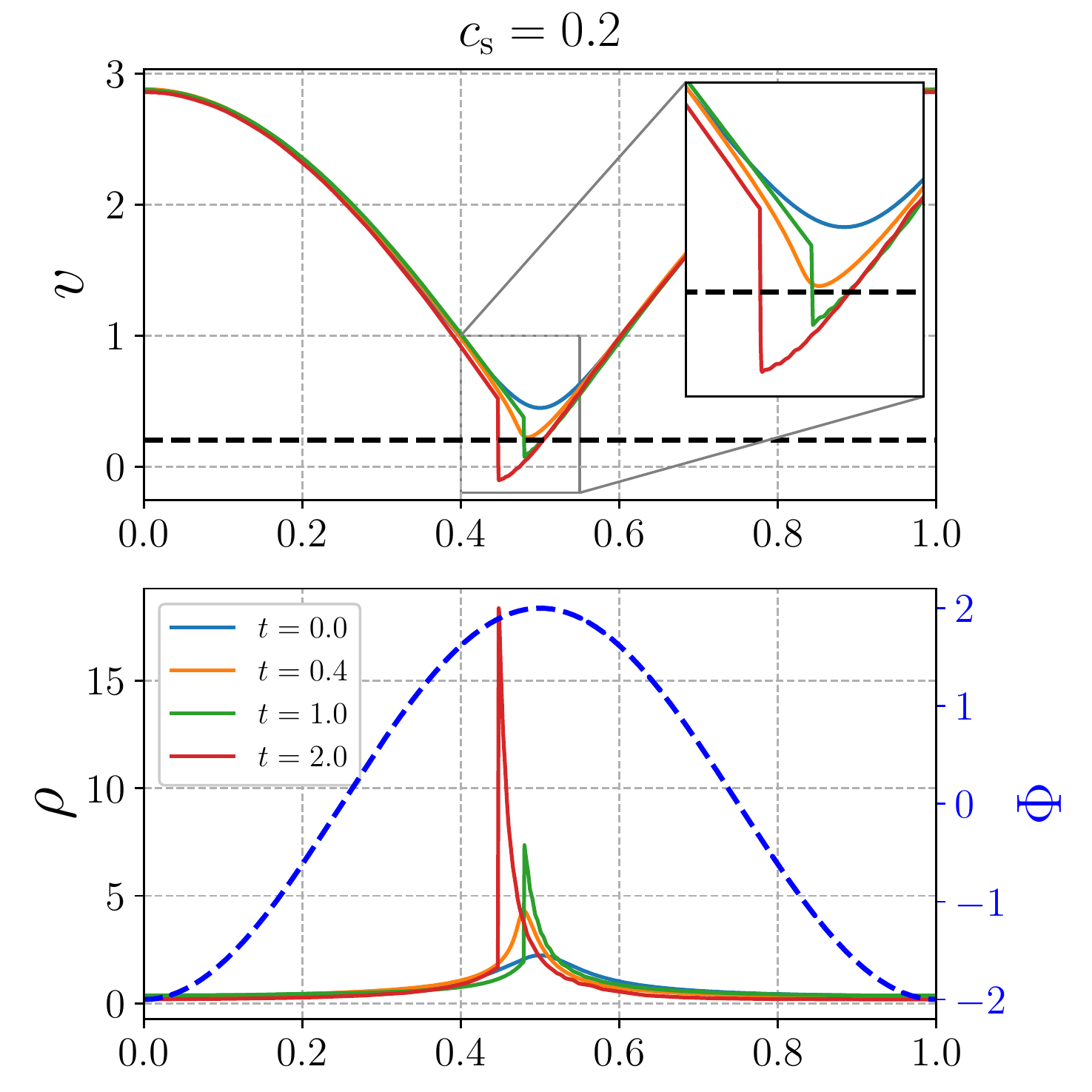}
\caption{Evolution of density and velocity in the 1D toy problem described in Section \ref{sec:pressure} for a high value of the sound speed, $\cs=0.2$. A shock is formed which moves upstream.}
\label{fig:tube_cs02}
\end{figure}

What happens when we introduce pressure? Suppose we release the system from an initial state given by Eqs. \eqref{eq:toysteady1} and \eqref{eq:toysteady2}. If $\cs=0$ nothing happens because the initial state is a steady state. But if $\cs>0$, the system is out of equilibrium and not exactly in a steady state. Fig. \ref{fig:tube_cs002} shows the subsequent evolution of the system if the initial state is the one with $B=2.1$ and the sound speed is $\cs=0.02$. In this case, pressure forces are negligible and the system remains approximately in the initial state. Fig. \ref{fig:tube_cs02} repeats the same experiment with a value of the sound speed $\cs=0.2$. Now, pressure forces are significant and quickly lead to the development of a shock (see Fig. \ref{fig:tube_cs02}). The shock initially develops close to the point where $v$ is minimum ($\Phi$ is maximum) and subsequently moves upstream. The reason why the shock develops is that as the gas tries to `climb up' from the potential well, pressure pushes backwards and slows it down, making the gas `stall' \citep{Prendergast1983}. The condition for the shock to develop is that the sound speed (black dashed line in Figs. \ref{fig:tube_cs002} and \ref{fig:tube_cs02}) is close enough to the minimum of $v$. Thus, the shocks occur if the sound speed is high enough (but also low enough that the gas is supersonic) to get close to the minimum of $v$. In a system with fixed $\cs$, a shock occurs if the oscillations of $\Phi$ are large enough.

Now we can understand what happens in the case of Fig. \ref{fig:tube}. The value $\cs=10\kms$ is close enough to the minimum value of $v$ along the $x_2$ orbit that a shock is triggered. Shocked gas is slowed down, and starts falling to the centre. This is why material in the spiral pattern moves \emph{inward}, despite the fact that the gravitational torques from the bar potential may suggest the opposite.\footnote{Indeed, it is known that spiral shocks can drive disc accretion \citep[e.g.][]{ArzamasskiyRafikov2018}.} After reaching its closest approach to the centre during its first infall, gas gains speed (the green curve in Fig. \ref{fig:tube} has a lower minimum and a higher maximum velocity than the x2 orbit velocity given by the blue curve). Thus while close to the shock gas is falling in, it is actually moving out far from the shock (dashed lines in Fig. \ref{fig:tube}). Meanwhile, the shock moves upstream, as in the toy model. This causes the overall gas distribution to tilt, causing the removal of angular momentum discussed in Sect. \ref{sec:results2}.

The shock in the simulations of Fig. \ref{fig:rho5} first appears in the outer parts, and then quickly propagates inwards. This happens because oscillations of $\Phi$ are stronger in the outer parts (see bottom-middle panel of Fig. \ref{fig:SBM_1}). As we move inwards, at some point the oscillations of $\Phi$ are not strong enough to trigger a shock, but merely a density wave. Thus, there is a critical radius $R_{\rm c}$ above which oscillations of $\Phi$ are strong enough to trigger a shock \citep{Roberts1969}. The conclusion is that the disc in Fig. \ref{fig:rho5} shrinks until it is small enough that shocks cannot be triggered anymore because the oscillations of $\Phi$ are small.

This also explains why if one increases the sound speed the nuclear disc shrinks, a trend which has been observed by many authors \citep[e.g.][]{englmaiergerhard1997,PatsisAthanassoula2000,Kim++2012a,SBM2015a}. When the sound speed is increased, smaller $\Phi$ oscillations are sufficient to trigger a shock (raising the black dashed in Figs. \ref{fig:tube_cs002} and \ref{fig:tube_cs02} favours the triggering of shocks). Hence, as the sound speed increases the value of $R_{\rm c}$ decreases, shocks can be produced further in and the disc shrinks further according to the considerations in the last paragraph. Other explanations of why the $x_2$ disc is smaller at higher sound speed also rely on shocks and the loss of angular momentum \citep{Kim++2012a,SBM2015a}. However, these explanation focus on the loss of angular momentum at the tips of the bar, far from the nuclear region. What is different and interesting in the explanation proposed here is that the part of the shock that matters in setting the size of the nuclear disc is that already within the $x_2$ region. Indeed, the simulation in Fig. \ref{fig:rho5} does not contain the full larger-scale off-set shock that typically develops in a gas flow in a barred potential, but just its last portion. We have confirmed this interpretation by repeating the simulation shown in Fig. \ref{fig:rho5} with different values of the sound speed. These show the same trend observed by other authors, i.e. the final size of the disc is smaller at higher sound speeds, but they do not contain the larger scale flow, only the $x_2$ disc. For example, when $\cs=5\kms$, spirals develop but the final disc is bigger than in the last panel of Fig. \ref{fig:rho5}. When $\cs=1\kms$, the pressure is too low to generate spiral shocks -- instead, a gap is created approximately between the green and orange region in Fig. \ref{fig:SBM_1} as expected based on the considerations in Sect. \ref{sec:theory}, and a ring is formed. 

\subsection{Gravitational torques vs viscous torques} \label{sec:gravtorq}

In the description of the mechanism of Sect. \ref{sec:theory} we have assumed that the time average of the net gravitational torque on the disc,
\begin{equation}
\left< \mathcal{T}_{\rm g} \right> = \int \left<  \rho \right> \left( \frac{\pa \Phi}{\pa \theta}\right)  \, \di x\,\di y ,
\end{equation}
can be considered small. This assumption is correct for a collection of ballistic particles on $x_2$ orbits, for which $\mathcal{T}_{\rm g}$ oscillates periodically in time, and for a setup such as the one described in Sect. \ref{sec:theory} and used as initial condition in the idealised simulations of Sect. \ref{sec:resultsvisc}, because the density distribution is symmetric with respect to the axis. However, is this assumption justified in a more realistic situation such as the one described in Sect. \ref{sec:chemresults}?

\begin{figure}
\centering
\includegraphics[width=0.48\textwidth]{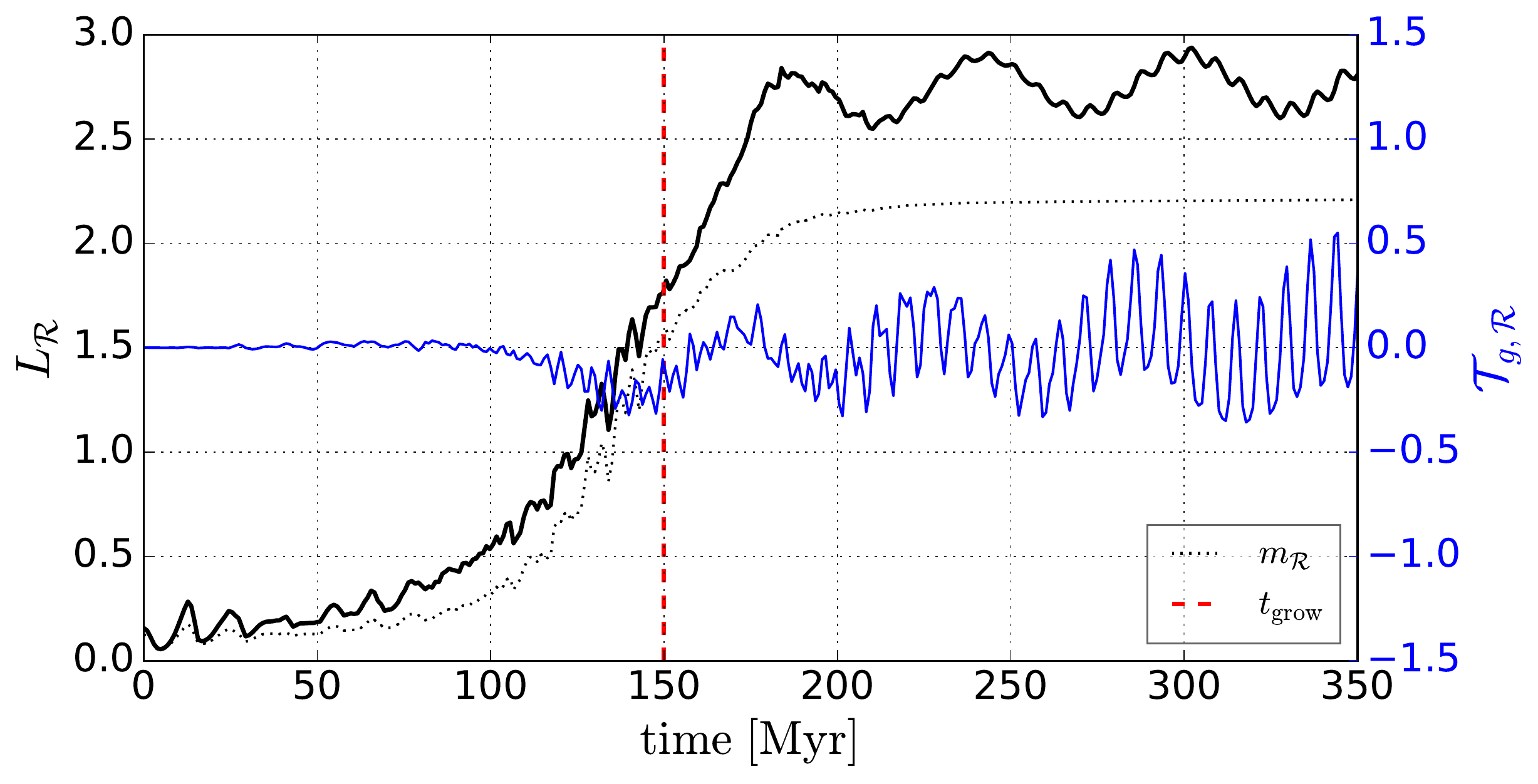}
\caption{\emph{Full black line:} total angular momentum contained in the region $\mathcal{R}$ (defined by Eq. \ref{eq:regionR}) of the `high' simulation of \protect\cite{Sormani2018}, as a function of time. \emph{Full blue line:} total gravitational torque exerted on material in the region $\mathcal{R}$. \emph{Dotted line:} total mass contained in the region $\mathcal{R}$ (values should be read on the left-side axis). The vertical red dashed line indicates the time at which the bar is fully grown. Units are as follows: $[L]= 10^8 M_\odot \ 100 \kms \kpc$, $[\mathcal{T}] = 10^8 M_\odot \ (100 \kms)^2$, $[m] = 3 \times 10^8 M_\odot$.}
\label{fig:torques_final}
\end{figure}

The black solid curve in Fig. \ref{fig:torques_final} shows the total angular momentum $L_{\mathcal{R}}$ possessed by gas in the simulations of \cite{Sormani2018} in the region defined as
\begin{equation} \label{eq:regionR}
\mathcal{R} : = \left\{ 0.2^2 < \left(\frac{x}{0.4\kpc}\right)^2 +  \left(\frac{y}{0.5 \kpc}\right)^2 < 1  \right\}
\end{equation}
i.e., a region comprised by two ellipses of semi-major axis of $0.1\kpc$ and $0.5\kpc$. This region roughly includes the ring shown in Fig. \ref{fig:chem}, while excluding outer material flowing on $x_1$ orbits and the innermost material. The blue solid line is the total gravitational torque $\mathcal{T}_{g,\mathcal{R}}$ acting on gas inside this region, and the dotted black curve  $m_{\mathcal{R}}$ is the total gas mass.

In the initial phase ($t \lesssim 180 \Myr$), the angular momentum in the region $\mathcal{R}$ increases steadily.\footnote{Recall that the bar is introduced gradually during the first $150 \Myr$ in the simulations of \cite{Sormani2018} in order to avoid transients.} This is mainly due to the fact that the nuclear region is accreting gas brought to $\mathcal{R}$ by the large-scale bar shocks (see the dotted line in Fig. \ref{fig:torques_final}), and not to the fact that the angular momentum per unit mass increases. After the initial phase, accretion almost stops, the mass in this region stays approximately constant and the angular momentum oscillates quasi-periodically. The gravitational torques oscillate and their time average is  clearly much lower than the amplitude of the oscillations.\footnote{The instantaneous gravitational torques are not zero because the gas distribution is not symmetric about the Galactic centre (showing this was one of the main goals of \citealt{Sormani2018}). Instead, the gas distribution is made up of a few giant molecular clouds that revolve around the centre approximately following $x_2$ orbits.} An estimate obtained by averaging the data in Fig. \ref{fig:torques_final} over a period $t = [200, 320] \Myr$, corresponding to approximately two of the longer oscillations, gives $ \langle L_{\mathcal{R}} \rangle / \langle \mathcal{T}_{g,\mathcal{R}} \rangle \simeq 3 \Gyr$. This is an estimate of the time scale over which gravitational torques have net effects in this region, and it is much longer than the duration of the simulation.

In the phase $t>180\Myr$, as can be seen from the movies linked above, the gas distribution is made of molecular clouds that approximately follow $x_2$ orbits (which act as guiding centres) on top of which there are significant excursions. A lot of turbulent motions are taking place, but over time these turbulent motions are dissipated, eventually leading to the narrow ring shown in Fig. \ref{fig:chem}. Thus, the gas does not really move in annuli smoothly exerting torques on each other as assumed in Sect. \ref{sec:theory}. However, it is likely that the formation of the ring does not depend too much on the details of the dissipation process (in this case, turbulent dissipation). The ring configuration is a local minimum of the energy, and the system can be driven to it regardless of the details of the dissipation process.

It is clearly seen by looking at the $L_{\mathcal{R}}$ curve in Fig. \ref{fig:torques_final} for $t>180\Myr$ that there are two main time scales of the oscillations, a short one with period $T_1\simeq 7 \Myr$ and a longer one with period $T_2\simeq60\Myr$. The smaller timescale $T_1$ corresponds to half a period of an $x_2$ orbit. Thus, these oscillations resemble those that we would obtain if we had a bunch of particles flowing on $x_2$ orbits. The longer time-scale $T_2$ corresponds to a very interesting phenomenon, that is the major axis of the $x_2$ disc/ring oscillates so that its angle with the $y$ axis changes periodically. 

The picture that emerges from the above considerations is one in which the gravitational torques are important in bringing down gas from from the outer parts down to the nuclear region (with the mediation of shocks). Once material is in the nuclear region, it settles on $x_2$ orbits and from that point viscosity becomes important and a ring of the type described in Sect. \ref{sec:theory} is developed.

\subsection{Relevance for real galaxies} \label{sec:realgalaxies}

\subsubsection{Effective equation of state of the ISM}

The results of Sect. \ref{sec:resultsvisc} and \ref{sec:results2} (see also the discussions of Sect. \ref{sec:pressure} and \ref{sec:gravtorq}) suggest that the mechanism described in Sect. \ref{sec:theory} is a viable way of ring formation if the gas sound speed is low ($\cs \lesssim 1\kms$), but not if the gas sound speed is high ($\cs \simeq 10 \kms$). In the chemistry simulations of Sect. \ref{sec:chemresults} the mechanism is able to work because the cold phase of the two-phase medium effectively behaves as a low sound speed gas. Thus, whether our theory can be applied to real galaxies depends on the `effective' equation of state of the interstellar medium (ISM).

What is the equation of state of the ISM? This is a very difficult question because the ISM is a complex dynamical entity and it is not easy to justify a set of equations that capture its dynamics. It is made of several components that are tightly coupled together. The four most important of these, in terms of their impact on the overall dynamical evolution, are gas, dust, cosmic rays, and magnetic fields. In addition, the ISM is highly inhomogeneous and shows structure on all observable scales, with star formation occurring in dense molecular clouds, and it is subject to several complex feedback loops driven by the energy and momentum input from stars in form of radiation, winds, and supernovae.  

There are two very different approaches to this complexity in the literature. The first one is to simply ignore it. If one looks at the properties of the atomic component of the ISM, as revealed by its $21\,$cm emission, then one finds a remarkably constant velocity dispersion in the range $5\, \mhyphen \,20 \kms$ quite independent of galactic radius and galaxy type \cite[e.g.][]{Walter08a}. This has lead many studies of ISM dynamics on galactic scales to adopt an  `isothermal prescription' and to take the gas to be isothermal with a sound speed of  $\cs \simeq 10 \kms$, where the sound speed is \emph{not} related to the actual kinetic temperature of the gas, but corresponds to an `effective' temperature which is supposed to account for the turbulent motions in the ISM on unresolved scales \citep[e.g.][]{Roberts1969,Cowie1980,Athan92b,englmaiergerhard1997,ReganTeuben2003,Maciejewski2004b,li05, li06, Kim++2012a,SBM2015a,Fragkoudi+2017}.  The mechanism of ring formation presented in this paper generally does not apply if the ISM is modelled using this isothermal prescription, because the sound speed typically assumed ($\cs \simeq 5\, \mhyphen \,20 \kms$) is too high.\footnote{Although it may apply to some simulations which assumed low sound speeds, for example the lowest sound speed simulations of \cite{PatsisAthanassoula2000} and \cite{Kim++2012a} which form a ring.} However, as discussed above this treatment of the ISM is very crude. A similar simplification comes from the use of sticky-particle techniques, which model the gas as an ensemble of inelastically colliding particles \citep[e.g.][]{Schwarz1981,CombesGerin1985,Byrd+1994,JenkinsBinney94,RautiainenSalo2000,Rautiainen+2002,Rautiainen+2004,RFC2008}. The underlying assumption is that the ISM is mostly composed of long-lived dense molecular clouds, that are essentially described as `bricks', that only interact collisionally \cite[e.g.][]{Tan00}. Clearly this approach does not do justice to the  multi-phase nature and chemical complexity of the observed ISM \cite[][]{KlessenGlover2016}.

The second approach is to face the observed complexity and construct a model that takes into account as many physical and chemical processes as possible and also includes a realistic treatment of stellar feedback. With the advent of new numerical modeling techniques and the continuously increasing capabilities provided by modern supercomputing centers this approach has become very popular in recent years. This has started with detailed modelling of a small portion of the galaxy or an individual molecular cloud, and we have now reached the point where it is feasible to apply this detailed modelling to simulations of an entire galaxy  \cite[see e.g.][]{Smith+2014a, Bournaud15, Iffrig15, Walch15, Geen16, Girichidis16, Hu17, Kim17, Peters17, Seifried17, Safranek17, Hennebelle18, Hopkins18, Pillepich18}. The simulations of \cite{Sormani2018} follow this tradition and arguably provide a more realistic treatment of the ISM than the isothermal prescription, and the mechanism presented in this paper seems to work in this case. The is probably due to the fact that indeed a large fraction of the mass in the nuclear region (called the Central Molecular Zone or CMZ in the Milky Way) is relatively cold with a sound speed smaller than $1 \kms$. 

This is a reassuring confirmation, but the approach of \cite{Sormani2018} still lacks several important ingredients. Most importantly, it lacks the small scale turbulence of the ISM, which is believed to be driven by supernova explosions and other stellar feedback mechanisms \citep[e.g.][]{KlessenGlover2016}. This is evident in the fact that the ring in Fig. \ref{fig:chem} is too thin compared to real rings. It also does not include magnetic fields, which may also affect the dynamics significantly. A useful test would be to run simulations similar to those of \cite{Sormani2018} but including the effects of stellar feedback, to see whether the size of the ring is affected or not. If the results of the chemistry simulations are confirmed, it would support the idea that the mechanism is relevant to real galaxies.

We conclude that whether or not the mechanism presented in this paper is relevant for real galaxies depends on the nature of the effective equation of state of the ISM. The question is complex and not settled yet, but future simulations may provide answers.

\subsubsection{Morphology of nuclear regions}

The discussion of Sect. \ref{sec:pressure} suggests that the morphology of the nuclear region depends on the effective sound speed of the gas, i.e.\ that it depends on the importance of pressure forces. Lower pressure leads to more ring-like morphologies, while higher pressure leads to more spiral-like morphologies. The effective pressure of the ISM in real galaxies might be related to different temperature regimes of a multi-phase medium, but may also contain contributions from turbulent pressure (hence related to the velocity dispersion of clouds) and magnetic pressure. This may provide insight into observations of external galaxies where the nuclear region displays different morphologies in different tracers (\citealt{Izumi+2013,Fathi+2013,Izumi+2015,Espada+2017}, see also \citealt{Burillo+2005}). Different morphologies may reflect different temperature regimes of a multi-phase medium, or different magnetic field strengths.

\subsubsection{Constraining the pattern speed}

The pattern speed $\Omegap$ is the single most important parameter of a bar, but is notoriously difficult to measure \citep[e.g.][]{TremaineWeinberg1984,Rautiainen+2008,Aguerri+2015}. Thus methods that allow to constrain it are very valuable. The size of the ring in our theory is sensitive to the underlying gravitational potential. In particular, it is sensitive to the pattern speed of the bar: a higher pattern speed produces a smaller ring as it shrinks the extent of the $x_2$ family of periodic orbits. Hence if the mass distribution of a barred potential is known but the pattern speed is not, our theory can be used to constrain $\Omegap$ by adjusting its value until the reversed shear region matches the observed size of the ring. The size of the ring predicted by our theory also depends on other properties of the potentials (for example a stronger bar generally produces smaller rings), which could therefore also be constrained.

\subsubsection{Accretion onto the central black hole}

The ring configuration is a configuration of `local minimum' of the energy. Hence, to get out of this ring and reach the \emph{global} minimum of the energy (which corresponds to most of the mass in the centre) gas needs to overcome an `energy barrier'. The question of how the gas may overcome this energy barrier might be related to how the gas migrates from the nuclear ring ($r \sim 100 \pc$) down to an accretion disc that fuels a supermassive black hole ($r \sim 10^{-3} \pc$). Since nuclear rings are sites of intense star formation, we might speculate that stellar feedback `launches' parcels of gas that overcome the barrier and can then plunge towards the black hole \citep{Davies+2007}.

\subsection{Relation to other works} \label{sec:others}

\subsubsection{Galactic dynamics}

As we mentioned in the introduction the main theory for the formation of nuclear rings is the resonant theory, which suggests that they are related to the ILR of the underlying gravitational potential \citep[e.g.][]{Combes1988,Combes1996b,ButaCombes1996}. In our model the size of the ring is also indirectly related to the ILR, as it is well known that the radius of the ILR correlates with the radial extent of the $x_2$ orbit family \citep{contopoulosreview,Athan92a}, and therefore with the size of the reversed shear region (see Sect. \ref{sec:theory}). The main difference between our model and the resonant theory is that the latter relies only on gravitational torques, while our model also relies on viscous torques (see the discussion in Sect. \ref{sec:gravtorq}). In Sect. \ref{sec:results2} and \ref{sec:pressure} we have also argued that, when examined in more detail, the predictions of the resonant theory seem not to be verified in simulations because naive application of the idea that torques act on the spiral pattern may lead to incorrect conclusions.

An alternative theory based on viscous torques was proposed by \cite{Icke1979} and \cite{Lesch+1990} and recently revisited by \cite{KrumholzKruijssen2015}. This theory is based on the idea that in a circular accretion disc (with $\di \Omega/\di R < 0$) gas accumulates at the point where $\Omega(R)$ becomes flat (as in solid body rotation) because this reduces the shear responsible for the inward mass transfer. The predictions from this theory are different from the predictions of the mechanism proposed in Sect. \ref{sec:theory}. First, the location where the rotation curve turns solid body is in general very different from the location of the reversed shear region: as can be seen from Eq. \eqref{eq:log} the rotation curve in Fig. \ref{fig:SBM_2} only turns to solid body around $R_c = 0.05 \kpc$, while the reversed shear region is at $R \sim 1.2\kpc$. Second, according to such theories the size of the ring does not depend on the pattern speed of the bar, in contrast to what is predicted by the mechanism of Sect. \ref{sec:theory} and what is observed in simulations \citep[e.g.][]{Athan92b,Li+2015,SBM2015c}. For similar reasons, this theory cannot be the explanation for the ring that forms in the simulations of \cite{Sormani2018}: the ring does not form near the point at which the rotation curve turns solid body.

\cite{Li+2015} (see also \citealt{Kim++2012a}) ran various two-dimensional isothermal simulations of gas flow in barred potentials with varying bar strengths and pattern speeds. These authors use the `isothermal prescription' (see Sect. \ref{sec:realgalaxies}), hence they are in the high sound speed regime in which our mechanism does not apply (see Sect. \ref{sec:pressure}). However, they argue that the size of the ring is determined by the amount of angular momentum lost as the gas enters and travels along the bar shocks, which is different from the interpretation we have given in Sect. \ref{sec:pressure}. They also argue that a stronger bar removes more angular momentum, leading to smaller rings. The same correlation between bar strength and size of the ring is expected in our theory, but for a different reason: a stronger bar changes the orbital structure of the potential, which results in smaller $x_2$ orbits \citep[e.g.][]{contopoulosreview,Athan92a}. This correlation was also observed in the sticky-particle simulations of \cite{Salo+1999} and is consistent with the observational result that `stronger bars host smaller rings' \citep{Comeron2010}.

\cite{Binney++1991} (see also \citealt{SBM2015a}) predicts that the major axis of the ring should be equal to the minor axis of the cusped $x_1$ orbit. This predicts a ring whose size is generally significantly smaller than that of Sect. \ref{sec:theory}, although the details depend on the gravitational potential. Indeed, it would predict a ring whose size is much smaller than the one developed in the simulations of \cite{Sormani2018}. Note also that Fig. \ref{fig:chem} suggests that the bar shocks touch the ring near the point of its maximum extension in the $y$ direction, but tests we have performed with different potentials suggest that this is not generally the case: sometimes the shock can intersect the ring at lower or values of $y$, and sometimes it can intersect the $y$ axis further out, flying by the ring without touching it. This suggests that the point where the shock touches the ring is not what determines the size of the ring.

\subsubsection{Planetary ring dynamics} \label{sec:planetary}

The problem studied in this paper has many analogies with the confinement of narrow rings in the context of planetary ring dynamics (see for example \citealt{Longaretti2018} for a recent review). In particular, the idea of reverse shear discussed in Sect. \ref{sec:theory} is equivalent to the idea of angular momentum flux reversal that has been developed by Borderies-Goldreich-Tremaine in a series of papers (e.g. \citealt{Borderies+1982,Borderies+1983a,Borderies+1983b,Borderies+1989,Borderies1989}) to explain the confinement and the sharp edges of planetary rings. These authors showed that under the influence of a perturbing potential, created for example by a satellite, the particle streamlines can be sufficiently distorted so that the angular momentum flux is directed inwards, which is very similar to the picture we described in Sect. \ref{sec:theory} (in our notation, the angular momentum flux is directed inwards when $k<0$). In the context of planetary rings this idea has been explored also using N-body simulations, both global (\citealt{HanninenSalo1994,HanninenSalo1995}; see also \citealt{Goldreich+1995} for a theoretical discussion of their results) and limited to a local comoving region \citep{Mosqueira1996}.

An obvious difference between the galactic and the planetary problems is the order of magnitude of the perturbations. While for a strong barred potential such as the one considered in Sect. \ref{sec:theory} particle orbits deviate strongly from circular motions (see Fig. \ref{fig:SBM_1}), in the case of planetary ring dynamics the deviations from circularity are typically very small, with eccentricities of the order of $10^{-5}$ \citep[e.g.][]{Borderies1989}. This reflects the fact that the resonance zone, i.e. the region where the effect of the resonances is non negligible, is much narrower in the planet-satellite case. Another major difference between the two problems is the mechanism which generates viscosity, namely collisions between particles in the planetary case and turbulent viscosity in the galactic case.

\section{Summary} \label{sec:summary}

We have developed a dynamical theory to explain why nuclear rings arise and to predict their location given the underlying gravitational potential. The main conclusions of this paper are as follows:

\begin{enumerate}
\item When the gas follows non-circular orbits such as $x_2$ orbits, viscous torques between adjacent annuli of gas can drive the gas towards a ring configuration (see Sect. \ref{sec:theory}), in contrast to what is found in the elementary theory of circular viscous accretion discs in which all the gas is driven to the centre. The ring configuration is a local minimum of the energy.
\item The sound speed and the amount of viscosity are two key parameters of the gas flow in the nuclear regions of galaxies.
\item When the sound speed is low ($\cs \lesssim 1 \kms$), a ring may be formed according to the mechanism of item (i). The location of this ring correlates with the ILR and can be approximately predicted based on the underlying gravitational potential: it must be within the reversed shear region (orange in Fig. \ref{fig:SBM_1}). We have speculated that the ring size might be predicted even more precisely as the orbit where $K_\parallel=0$, but this requires further verification (see Sect. \ref{sec:predictions}).
\item When the sound speed is high ($\cs \simeq 10 \kms$), such a ring cannot survive because spiral shocks are developed, regardless of the presence of viscosity. In this case, the size of the nuclear region is determined by the critical radius $R_c$ at which the oscillations of the gravitational potential become strong enough to induce a shock. The value of $R_c$ decreases at higher sound speeds, leading to smaller nuclear regions, according to the discussion in Sect. \ref{sec:pressure}.
\item Whether the mechanism described in Sect. \ref{sec:theory} is relevant for the formation of nuclear rings in real galaxies depends on the `effective' equation of state of the ISM. This is a complex question that is not settled yet, but simulations will provide great help in the near future.
\end{enumerate}

Putting everything together, we argued that a plausible scenario of ring formation (see Sect. \ref{sec:gravtorq}) in a real galaxy involves two phases: in the first phase gravitational torques bring the gas from outside the bar region down to the nuclear region. Then, once gas is within the nuclear region, gravitational torques become less important and turbulent dissipation while it settles on $x_2$ orbits will drive the gas towards the stable ring configuration described in Sect. \ref{sec:theory}.

\section*{Acknowledgements}

The authors thank the anonymous referee for pointing out the connection between the present work and the work by Borderies-Goldreich-Tremaine in the context of planetary ring dynamics. The authors also thank James Binney, David Cole, Payel Das, Cornelis Dullemond, Sacha Hony, Sarah Jeffreson, Zhi Li, Yuri Lyubarsky, Diederik Kruijssen, Mattis Magg, John Magorrian, Eva Schinnerer and Ralph Schoenrich for useful comments and discussions. MCS thanks Andrea Mignone and Claudio Zanni for help with the {\sc Pluto} code. MCS, RGT, SCOG, and RSK acknowledge support from the Deutsche Forschungsgemeinschaft via the Collaborative Research Centre (SFB 881)
``The Milky Way System'' (sub-projects B1, B2, and B8) and the Priority Program SPP 1573 ``Physics of the Interstellar Medium'' (grant numbers KL 1358/18.1, KL 1358/19.2, and GL 668/2-1). RSK furthermore thanks the European Research Council for funding in  the ERC Advanced Grant STARLIGHT (project number 339177). ES acknowledges support from the Israeli Science Foundation under Grant No. 719/14. The simulations in this paper were run using the bwForCluster MLS\&WISO; the authors acknowledge support by the state of Baden-W\"urttemberg through bwHPC and the German Research Foundation (DFG) through grant INST 35/1134-1 FUGG.

\def\aap{A\&A}\def\aj{AJ}\def\apj{ApJ}\def\mnras{MNRAS}\def\araa{ARA\&A}\def\aapr{Astronomy \&
 Astrophysics Review}\def\apjs{ApJS}\def\apjl{ApJ}\def\pasj{PASJ}\def\nat{Nature}\def\prd{Phys. Rev. D}
\def\ssr{Space Sci. Rev.}\def\pasp{PASP}\def\icarus{Icarus}
\bibliographystyle{mn2e}
\bibliography{bibliography}

\begin{thebibliography}{99}
\expandafter\ifx\csname natexlab\endcsname\relax\def\natexlab#1{#1}\fi

\bibitem[{{Aguerri} {et~al}\mbox{.}(2015){Aguerri}, {M{\'e}ndez-Abreu},
  {Falc{\'o}n-Barroso}, {Amorin}, {Barrera-Ballesteros}, {Cid Fernandes},
  {Garc{\'{\i}}a-Benito}, {Garc{\'{\i}}a-Lorenzo}, {Gonz{\'a}lez Delgado},
  {Husemann}, {Kalinova}, {Lyubenova}, {Marino}, {M{\'a}rquez}, {Mast},
  {P{\'e}rez}, {S{\'a}nchez}, {van de Ven}, {Walcher}, {Backsmann},
  {Cortijo-Ferrero}, {Bland-Hawthorn}, {del Olmo}, {Iglesias-P{\'a}ramo},
  {P{\'e}rez}, {S{\'a}nchez-Bl{\'a}zquez}, {Wisotzki}, \&
  {Ziegler}}]{Aguerri+2015}
{Aguerri} J.~A.~L. {et~al.}, 2015, \aap, 576, A102

\bibitem[{{Arzamasskiy} \& {Rafikov}(2018)}]{ArzamasskiyRafikov2018}
{Arzamasskiy} L., {Rafikov} R.~R., 2018, \apj, 854, 84

\bibitem[{{Athanassoula}(1992{\natexlab{a}})}]{Athan92a}
{Athanassoula} E., 1992{\natexlab{a}}, \mnras, 259, 328

\bibitem[{{Athanassoula}(1992{\natexlab{b}})}]{Athan92b}
{Athanassoula} E., 1992{\natexlab{b}}, \mnras, 259, 345

\bibitem[{{Binney} {et~al}\mbox{.}(1991){Binney}, {Gerhard}, {Stark}, {Bally},
  \& {Uchida}}]{Binney++1991}
{Binney} J., {Gerhard} O.~E., {Stark} A.~A., {Bally} J., {Uchida} K.~I., 1991,
  \mnras, 252, 210

\bibitem[{{Binney} \& {Tremaine}(2008)}]{BT2008}
{Binney} J., {Tremaine} S., 2008, {Galactic Dynamics: Second Edition}.
  Princeton University Press

\bibitem[{{Borderies}(1989)}]{Borderies1989}
{Borderies} N., 1989, Celestial Mechanics and Dynamical Astronomy, 46, 207

\bibitem[{{Borderies} {et~al}\mbox{.}(1982){Borderies}, {Goldreich}, \&
  {Tremaine}}]{Borderies+1982}
{Borderies} N., {Goldreich} P., {Tremaine} S., 1982, \nat, 299, 209

\bibitem[{{Borderies} {et~al}\mbox{.}(1983{\natexlab{a}}){Borderies},
  {Goldreich}, \& {Tremaine}}]{Borderies+1983a}
{Borderies} N., {Goldreich} P., {Tremaine} S., 1983{\natexlab{a}}, \aj, 88, 226

\bibitem[{{Borderies} {et~al}\mbox{.}(1983{\natexlab{b}}){Borderies},
  {Goldreich}, \& {Tremaine}}]{Borderies+1983b}
{Borderies} N., {Goldreich} P., {Tremaine} S., 1983{\natexlab{b}}, \aj, 88,
  1560

\bibitem[{{Borderies} {et~al}\mbox{.}(1989){Borderies}, {Goldreich}, \&
  {Tremaine}}]{Borderies+1989}
{Borderies} N., {Goldreich} P., {Tremaine} S., 1989, \icarus, 80, 344

\bibitem[{{Bournaud} {et~al}\mbox{.}(2015){Bournaud}, {Daddi}, {Wei{\ss}},
  {Renaud}, {Mastropietro}, \& {Teyssier}}]{Bournaud15}
{Bournaud} F., {Daddi} E., {Wei{\ss}} A., {Renaud} F., {Mastropietro} C.,
  {Teyssier} R., 2015, \aap, 575, A56

\bibitem[{{Buta} \& {Combes}(1996)}]{ButaCombes1996}
{Buta} R., {Combes} F., 1996, Fund. Cosmic Physics, 17, 95

\bibitem[{{Buta}(2017{\natexlab{a}})}]{Buta2017a}
{Buta} R.~J., 2017{\natexlab{a}}, \mnras, 471, 4027

\bibitem[{{Buta}(2017{\natexlab{b}})}]{Buta2017b}
{Buta} R.~J., 2017{\natexlab{b}}, \mnras, 470, 3819

\bibitem[{{Byrd} {et~al}\mbox{.}(1994){Byrd}, {Rautiainen}, {Salo}, {Buta}, \&
  {Crocher}}]{Byrd+1994}
{Byrd} G., {Rautiainen} P., {Salo} H., {Buta} R., {Crocher} D.~A., 1994, \aj,
  108, 476

\bibitem[{{Combes}(1988)}]{Combes1988}
{Combes} F., 1988, in NATO Advanced Science Institutes (ASI) Series C, Vol.
  232, NATO Advanced Science Institutes (ASI) Series C, {Pudritz} R.~E., {Fich}
  M., eds., p. 475

\bibitem[{{Combes}(1996)}]{Combes1996b}
{Combes} F., 1996, in Astronomical Society of the Pacific Conference Series,
  Vol.~91, IAU Colloq. 157: Barred Galaxies, {Buta} R., {Crocker} D.~A.,
  {Elmegreen} B.~G., eds., p. 286

\bibitem[{{Combes}(2001)}]{Combes2001}
{Combes} F., 2001, in Advanced Lectures on the Starburst-AGN, {Aretxaga} I.,
  {Kunth} D., {M{\'u}jica} R., eds., p. 223

\bibitem[{{Combes} \& {Gerin}(1985)}]{CombesGerin1985}
{Combes} F., {Gerin} M., 1985, \aap, 150, 327

\bibitem[{{Comer{\'o}n}(2013)}]{Comeron2013}
{Comer{\'o}n} S., 2013, \aap, 555, L4

\bibitem[{{Comer{\'o}n} {et~al}\mbox{.}(2010){Comer{\'o}n}, {Knapen},
  {Beckman}, {Laurikainen}, {Salo}, {Mart{\'{\i}}nez-Valpuesta}, \&
  {Buta}}]{Comeron2010}
{Comer{\'o}n} S., {Knapen} J.~H., {Beckman} J.~E., {Laurikainen} E., {Salo} H.,
  {Mart{\'{\i}}nez-Valpuesta} I., {Buta} R.~J., 2010, \mnras, 402, 2462

\bibitem[{{Contopoulos} \& {Grosbol}(1989)}]{contopoulosreview}
{Contopoulos} G., {Grosbol} P., 1989, \aapr, 1, 261

\bibitem[{{Cowie}(1980)}]{Cowie1980}
{Cowie} L.~L., 1980, \apj, 236, 868

\bibitem[{{Davies} {et~al}\mbox{.}(2007){Davies}, {M{\"u}ller S{\'a}nchez},
  {Genzel}, {Tacconi}, {Hicks}, {Friedrich}, \& {Sternberg}}]{Davies+2007}
{Davies} R.~I., {M{\"u}ller S{\'a}nchez} F., {Genzel} R., {Tacconi} L.~J.,
  {Hicks} E.~K.~S., {Friedrich} S., {Sternberg} A., 2007, \apj, 671, 1388

\bibitem[{{Englmaier} \& {Gerhard}(1997)}]{englmaiergerhard1997}
{Englmaier} P., {Gerhard} O., 1997, \mnras, 287, 57

\bibitem[{{Espada} {et~al}\mbox{.}(2017){Espada}, {Matsushita}, {Miura},
  {Israel}, {Neumayer}, {Martin}, {Henkel}, {Izumi}, {Iono}, {Aalto}, {Ott},
  {Peck}, {Quillen}, \& {Kohno}}]{Espada+2017}
{Espada} D. {et~al.}, 2017, \apj, 843, 136

\bibitem[{{Fathi} {et~al}\mbox{.}(2013){Fathi}, {Lundgren}, {Kohno},
  {Pi{\~n}ol-Ferrer}, {Mart{\'{\i}}n}, {Espada}, {Hatziminaoglou}, {Imanishi},
  {Izumi}, {Krips}, {Matsushita}, {Meier}, {Nakai}, {Sheth}, {Turner}, {van de
  Ven}, \& {Wiklind}}]{Fathi+2013}
{Fathi} K. {et~al.}, 2013, \apjl, 770, L27

\bibitem[{{Fragkoudi} {et~al}\mbox{.}(2017){Fragkoudi}, {Athanassoula}, \&
  {Bosma}}]{Fragkoudi+2017}
{Fragkoudi} F., {Athanassoula} E., {Bosma} A., 2017, \mnras, 466, 474

\bibitem[{{Garc\'ia-Burillo} {et~al}\mbox{.}(2005){Garc\'ia-Burillo}, {Combes},
  {Schinnerer}, {Boone}, \& {Hunt}}]{Burillo+2005}
{Garc\'ia-Burillo} S., {Combes} F., {Schinnerer} E., {Boone} F., {Hunt} L.~K.,
  2005, \aap, 441, 1011

\bibitem[{{Geen} {et~al}\mbox{.}(2016){Geen}, {Hennebelle}, {Tremblin}, \&
  {Rosdahl}}]{Geen16}
{Geen} S., {Hennebelle} P., {Tremblin} P., {Rosdahl} J., 2016, \mnras, 463,
  3129

\bibitem[{{Girichidis} {et~al}\mbox{.}(2016){Girichidis}, {Naab}, {Walch},
  {Hanasz}, {Mac Low}, {Ostriker}, {Gatto}, {Peters}, {W{\"u}nsch}, {Glover},
  {Klessen}, {Clark}, \& {Baczynski}}]{Girichidis16}
{Girichidis} P. {et~al.}, 2016, \apjl, 816, L19

\bibitem[{{Goldreich} {et~al}\mbox{.}(1995){Goldreich}, {Rappaport}, \&
  {Sicardy}}]{Goldreich+1995}
{Goldreich} P., {Rappaport} N., {Sicardy} B., 1995, \icarus, 118, 414

\bibitem[{{H{\"a}nninen} \& {Salo}(1994)}]{HanninenSalo1994}
{H{\"a}nninen} J., {Salo} H., 1994, \icarus, 108, 325

\bibitem[{{H{\"a}nninen} \& {Salo}(1995)}]{HanninenSalo1995}
{H{\"a}nninen} J., {Salo} H., 1995, \icarus, 117, 435

\bibitem[{{Hennebelle}(2018)}]{Hennebelle18}
{Hennebelle} P., 2018, \aap, 611, A24

\bibitem[{{Henshaw} {et~al}\mbox{.}(2016){Henshaw}, {Longmore}, {Kruijssen},
  {Davies}, {Bally}, {Barnes}, {Battersby}, {Burton}, {Cunningham}, {Dale},
  {Ginsburg}, {Immer}, {Jones}, {Kendrew}, {Mills}, {Molinari}, {Moore}, {Ott},
  {Pillai}, {Rathborne}, {Schilke}, {Schmiedeke}, {Testi}, {Walker}, {Walsh},
  \& {Zhang}}]{Henshaw+16a}
{Henshaw} J.~D. {et~al.}, 2016, \mnras, 457, 2675

\bibitem[{{Heyer} \& {Dame}(2015)}]{HeyerDame2015}
{Heyer} M., {Dame} T.~M., 2015, \araa, 53, 583

\bibitem[{{Hopkins} {et~al}\mbox{.}(2018){Hopkins}, {Wetzel}, {Kere{\v s}},
  {Faucher-Gigu{\`e}re}, {Quataert}, {Boylan-Kolchin}, {Murray}, {Hayward}, \&
  {El-Badry}}]{Hopkins18}
{Hopkins} P.~F. {et~al.}, 2018, \mnras, 477, 1578

\bibitem[{{Hu} {et~al}\mbox{.}(2017){Hu}, {Naab}, {Glover}, {Walch}, \&
  {Clark}}]{Hu17}
{Hu} C.-Y., {Naab} T., {Glover} S.~C.~O., {Walch} S., {Clark} P.~C., 2017,
  \mnras, 471, 2151

\bibitem[{{Icke}(1979)}]{Icke1979}
{Icke} V., 1979, \aap, 78, 21

\bibitem[{{Iffrig} \& {Hennebelle}(2015)}]{Iffrig15}
{Iffrig} O., {Hennebelle} P., 2015, \aap, 576, A95

\bibitem[{{Izumi} {et~al}\mbox{.}(2015){Izumi}, {Kohno}, {Aalto}, {Doi},
  {Espada}, {Fathi}, {Harada}, {Hatsukade}, {Hattori}, {Hsieh}, {Ikarashi},
  {Imanishi}, {Iono}, {Ishizuki}, {Krips}, {Mart{\'{\i}}n}, {Matsushita},
  {Meier}, {Nagai}, {Nakai}, {Nakajima}, {Nakanishi}, {Nomura}, {Regan},
  {Schinnerer}, {Sheth}, {Takano}, {Tamura}, {Terashima}, {Tosaki}, {Turner},
  {Umehata}, \& {Wiklind}}]{Izumi+2015}
{Izumi} T. {et~al.}, 2015, \apj, 811, 39

\bibitem[{{Izumi} {et~al}\mbox{.}(2013){Izumi}, {Kohno}, {Mart{\'{\i}}n},
  {Espada}, {Harada}, {Matsushita}, {Hsieh}, {Turner}, {Meier}, {Schinnerer},
  {Imanishi}, {Tamura}, {Curran}, {Doi}, {Fathi}, {Krips}, {Lundgren}, {Nakai},
  {Nakajima}, {Regan}, {Sheth}, {Takano}, {Taniguchi}, {Terashima}, {Tosaki},
  \& {Wiklind}}]{Izumi+2013}
{Izumi} T. {et~al.}, 2013, \pasj, 65, 100

\bibitem[{{Jenkins} \& {Binney}(1994)}]{JenkinsBinney94}
{Jenkins} A., {Binney} J., 1994, \mnras, 270, 703

\bibitem[{{Kim} \& {Ostriker}(2017)}]{Kim17}
{Kim} C.-G., {Ostriker} E.~C., 2017, \apj, 846, 133

\bibitem[{{Kim} {et~al}\mbox{.}(2014){Kim}, {Kim}, \& {Kim}}]{KimKimKim2014}
{Kim} W.-T., {Kim} Y., {Kim} J.-G., 2014, \apj, 789, 68

\bibitem[{{Kim} {et~al}\mbox{.}(2012){Kim}, {Seo}, {Stone}, {Yoon}, \&
  {Teuben}}]{Kim++2012a}
{Kim} W.-T., {Seo} W.-Y., {Stone} J.~M., {Yoon} D., {Teuben} P.~J., 2012, \apj,
  747, 60

\bibitem[{{Klessen} \& {Glover}(2016)}]{KlessenGlover2016}
{Klessen} R.~S., {Glover} S.~C.~O., 2016, in: Star Formation in Galaxy
  Evolution: Connecting Numerical Models to Reality, Saas-Fee Advanced Course,
  43, 85

\bibitem[{{Knapen}(2005)}]{Knapen2005}
{Knapen} J.~H., 2005, \aap, 429, 141

\bibitem[{{Kruijssen} {et~al}\mbox{.}(2015){Kruijssen}, {Dale}, \&
  {Longmore}}]{Kruijssen+2015}
{Kruijssen} J.~M.~D., {Dale} J.~E., {Longmore} S.~N., 2015, \mnras, 447, 1059

\bibitem[{{Krumholz} \& {Kruijssen}(2015)}]{KrumholzKruijssen2015}
{Krumholz} M.~R., {Kruijssen} J.~M.~D., 2015, \mnras, 453, 739

\bibitem[{{Lesch} {et~al}\mbox{.}(1990){Lesch}, {Biermann}, {Crusius},
  {Reuter}, {Dahlem}, {Barteldrees}, \& {Wielebinski}}]{Lesch+1990}
{Lesch} H., {Biermann} P.~L., {Crusius} A., {Reuter} H.~P., {Dahlem} M.,
  {Barteldrees} A., {Wielebinski} R., 1990, \mnras, 242, 194

\bibitem[{{Li} {et~al}\mbox{.}(2005){Li}, {Mac Low}, \& {Klessen}}]{li05}
{Li} Y., {Mac Low} M.-M., {Klessen} R.~S., 2005, \apj, 626, 823

\bibitem[{{Li} {et~al}\mbox{.}(2006){Li}, {Mac Low}, \& {Klessen}}]{li06}
{Li} Y., {Mac Low} M.-M., {Klessen} R.~S., 2006, \apj, 639, 879

\bibitem[{{Li} {et~al}\mbox{.}(2017){Li}, {Sellwood}, \& {Shen}}]{Li+2017}
{Li} Z., {Sellwood} J.~A., {Shen} J., 2017, \apj, 850, 67

\bibitem[{{Li} {et~al}\mbox{.}(2015){Li}, {Shen}, \& {Kim}}]{Li+2015}
{Li} Z., {Shen} J., {Kim} W.-T., 2015, \apj, 806, 150

\bibitem[{{Liszt}(2009)}]{Liszt2009}
{Liszt} H.~S., 2009, ArXiv e-prints 0905.1412

\bibitem[{Longaretti(2018)}]{Longaretti2018}
Longaretti P.-Y., 2018, Theory of Narrow Rings and Sharp Edges, in Planetary
  Ring Systems: Properties, Structure, and Evolution, Tiscareno, Matthew S. and
  Murray, Carl D.Editors, Cambridge Planetary Science, Cambridge University
  Press, p. 225

\bibitem[{{Lynden-Bell} \& {Kalnajs}(1972)}]{LyndenBellKalnajs1972}
{Lynden-Bell} D., {Kalnajs} A.~J., 1972, \mnras, 157, 1

\bibitem[{{Lynden-Bell} \& {Pringle}(1974)}]{LyndenBellPringle1974}
{Lynden-Bell} D., {Pringle} J.~E., 1974, \mnras, 168, 603

\bibitem[{{Mac Low} \& {Klessen}(2004)}]{MacLowKlessen2004}
{Mac Low} M.-M., {Klessen} R.~S., 2004, Reviews of Modern Physics, 76, 125

\bibitem[{{Maciejewski}(2004)}]{Maciejewski2004b}
{Maciejewski} W., 2004, \mnras, 354, 892

\bibitem[{{Mignone} {et~al}\mbox{.}(2007){Mignone}, {Bodo}, {Massaglia},
  {Matsakos}, {Tesileanu}, {Zanni}, \& {Ferrari}}]{Pluto2007}
{Mignone} A., {Bodo} G., {Massaglia} S., {Matsakos} T., {Tesileanu} O., {Zanni}
  C., {Ferrari} A., 2007, \apjs, 170, 228

\bibitem[{{Molinari} {et~al}\mbox{.}(2011){Molinari}, {Bally},
  {Noriega-Crespo}, {Compi{\`e}gne}, {Bernard}, {Paradis}, {Martin}, {Testi},
  {Barlow}, {Moore}, {Plume}, {Swinyard}, {Zavagno}, {Calzoletti}, {Di
  Giorgio}, {Elia}, {Faustini}, {Natoli}, {Pestalozzi}, {Pezzuto},
  {Piacentini}, {Polenta}, {Polychroni}, {Schisano}, {Traficante}, {Veneziani},
  {Battersby}, {Burton}, {Carey}, {Fukui}, {Li}, {Lord}, {Morgan}, {Motte},
  {Schuller}, {Stringfellow}, {Tan}, {Thompson}, {Ward-Thompson}, {White}, \&
  {Umana}}]{Molinari+2011}
{Molinari} S. {et~al.}, 2011, \apjl, 735, L33

\bibitem[{{Mosqueira}(1996)}]{Mosqueira1996}
{Mosqueira} I., 1996, \icarus, 122, 128

\bibitem[{{Patsis} \& {Athanassoula}(2000)}]{PatsisAthanassoula2000}
{Patsis} P.~A., {Athanassoula} E., 2000, \aap, 358, 45

\bibitem[{{Peters} {et~al}\mbox{.}(2017){Peters}, {Zhukovska}, {Naab},
  {Girichidis}, {Walch}, {Glover}, {Klessen}, {Clark}, \&
  {Seifried}}]{Peters17}
{Peters} T. {et~al.}, 2017, \mnras, 467, 4322

\bibitem[{{Phinney}(1994)}]{Phinney1994}
{Phinney} E.~S., 1994, in Mass-Transfer Induced Activity in Galaxies,
  {Shlosman} I., ed., p.~1

\bibitem[{{Pillepich} {et~al}\mbox{.}(2018){Pillepich}, {Springel}, {Nelson},
  {Genel}, {Naiman}, {Pakmor}, {Hernquist}, {Torrey}, {Vogelsberger},
  {Weinberger}, \& {Marinacci}}]{Pillepich18}
{Pillepich} A. {et~al.}, 2018, \mnras, 473, 4077

\bibitem[{{Prendergast}(1983)}]{Prendergast1983}
{Prendergast} K.~H., 1983, in IAU Symposium, Vol. 100, Internal Kinematics and
  Dynamics of Galaxies, {Athanassoula} E., ed., pp. 215--220

\bibitem[{{Pringle}(1981)}]{Pringle1981}
{Pringle} J.~E., 1981, \araa, 19, 137

\bibitem[{{Rautiainen} \& {Salo}(2000)}]{RautiainenSalo2000}
{Rautiainen} P., {Salo} H., 2000, \aap, 362, 465

\bibitem[{{Rautiainen} {et~al}\mbox{.}(2004){Rautiainen}, {Salo}, \&
  {Buta}}]{Rautiainen+2004}
{Rautiainen} P., {Salo} H., {Buta} R., 2004, \mnras, 349, 933

\bibitem[{{Rautiainen} {et~al}\mbox{.}(2002){Rautiainen}, {Salo}, \&
  {Laurikainen}}]{Rautiainen+2002}
{Rautiainen} P., {Salo} H., {Laurikainen} E., 2002, \mnras, 337, 1233

\bibitem[{{Rautiainen} {et~al}\mbox{.}(2008){Rautiainen}, {Salo}, \&
  {Laurikainen}}]{Rautiainen+2008}
{Rautiainen} P., {Salo} H., {Laurikainen} E., 2008, \mnras, 388, 1803

\bibitem[{{Regan} \& {Teuben}(2003)}]{ReganTeuben2003}
{Regan} M.~W., {Teuben} P., 2003, \apj, 582, 723

\bibitem[{{Roberts}(1969)}]{Roberts1969}
{Roberts} W.~W., 1969, \apj, 158, 123

\bibitem[{{Rodriguez-Fernandez} \& {Combes}(2008)}]{RFC2008}
{Rodriguez-Fernandez} N.~J., {Combes} F., 2008, \aap, 489, 115

\bibitem[{{Safranek-Shrader} {et~al}\mbox{.}(2017){Safranek-Shrader},
  {Krumholz}, {Kim}, {Ostriker}, {Klein}, {Li}, {McKee}, \&
  {Stone}}]{Safranek17}
{Safranek-Shrader} C., {Krumholz} M.~R., {Kim} C.-G., {Ostriker} E.~C., {Klein}
  R.~I., {Li} S., {McKee} C.~F., {Stone} J.~M., 2017, \mnras, 465, 885

\bibitem[{{Salo} {et~al}\mbox{.}(1999){Salo}, {Rautiainen}, {Buta}, {Purcell},
  {Cobb}, {Crocker}, \& {Laurikainen}}]{Salo+1999}
{Salo} H., {Rautiainen} P., {Buta} R., {Purcell} G.~B., {Cobb} M.~L., {Crocker}
  D.~A., {Laurikainen} E., 1999, \aj, 117, 792

\bibitem[{{Schwarz}(1981)}]{Schwarz1981}
{Schwarz} M.~P., 1981, \apj, 247, 77

\bibitem[{{Seifried} {et~al}\mbox{.}(2017){Seifried}, {Walch}, {Girichidis},
  {Naab}, {W{\"u}nsch}, {Klessen}, {Glover}, {Peters}, \& {Clark}}]{Seifried17}
{Seifried} D. {et~al.}, 2017, \mnras, 472, 4797

\bibitem[{{Sellwood} \& {Balbus}(1999)}]{SellwoodBalbus1999}
{Sellwood} J.~A., {Balbus} S.~A., 1999, \apj, 511, 660

\bibitem[{{Sellwood} \& {Wilkinson}(1993)}]{Sellwood1993}
{Sellwood} J.~A., {Wilkinson} A., 1993, Reports on Progress in Physics, 56, 173

\bibitem[{{Shakura} \& {Sunyaev}(1973)}]{ShakuraSunyaev1973}
{Shakura} N.~I., {Sunyaev} R.~A., 1973, \aap, 24, 337

\bibitem[{{Smith} {et~al}\mbox{.}(2014){Smith}, {Glover}, {Clark}, {Klessen},
  \& {Springel}}]{Smith+2014a}
{Smith} R.~J., {Glover} S.~C.~O., {Clark} P.~C., {Klessen} R.~S., {Springel}
  V., 2014, \mnras, 441, 1628

\bibitem[{{Sormani} {et~al}\mbox{.}(2015{\natexlab{a}}){Sormani}, {Binney}, \&
  {Magorrian}}]{SBM2015a}
{Sormani} M.~C., {Binney} J., {Magorrian} J., 2015{\natexlab{a}}, \mnras, 449,
  2421

\bibitem[{{Sormani} {et~al}\mbox{.}(2015{\natexlab{b}}){Sormani}, {Binney}, \&
  {Magorrian}}]{SBM2015b}
{Sormani} M.~C., {Binney} J., {Magorrian} J., 2015{\natexlab{b}}, \mnras, 451,
  3437

\bibitem[{{Sormani} {et~al}\mbox{.}(2015{\natexlab{c}}){Sormani}, {Binney}, \&
  {Magorrian}}]{SBM2015c}
{Sormani} M.~C., {Binney} J., {Magorrian} J., 2015{\natexlab{c}}, \mnras, 454,
  1818

\bibitem[{{Sormani} {et~al}\mbox{.}(2017){Sormani}, {Sobacchi}, {Shore},
  {Tre{\ss}}, \& {Klessen}}]{Sormani+2017}
{Sormani} M.~C., {Sobacchi} E., {Shore} S.~N., {Tre{\ss}} R.~G., {Klessen}
  R.~S., 2017, \mnras, 471, 2932

\bibitem[{{Sormani} {et~al}\mbox{.}(2018){Sormani}, {Tre{\ss}}, {Ridley},
  {Glover}, {Klessen}, {Binney}, {Magorrian}, \& {Smith}}]{Sormani2018}
{Sormani} M.~C., {Tre{\ss}} R.~G., {Ridley} M., {Glover} S.~C.~O., {Klessen}
  R.~S., {Binney} J., {Magorrian} J., {Smith} R., 2018, \mnras, 475, 2383

\bibitem[{{Springel}(2010)}]{Springel2010}
{Springel} V., 2010, \mnras, 401, 791

\bibitem[{{Tan}(2000)}]{Tan00}
{Tan} J.~C., 2000, \apj, 536, 173

\bibitem[{{Tremaine} \& {Weinberg}(1984)}]{TremaineWeinberg1984}
{Tremaine} S., {Weinberg} M.~D., 1984, \apjl, 282, L5

\bibitem[{{van de Ven} \& {Chang}(2009)}]{VandevenChang2009}
{van de Ven} G., {Chang} P., 2009, \apj, 697, 619

\bibitem[{{Wada} \& {Koda}(2004)}]{WadaKoda2004}
{Wada} K., {Koda} J., 2004, \mnras, 349, 270

\bibitem[{{Walch} {et~al}\mbox{.}(2015){Walch}, {Girichidis}, {Naab}, {Gatto},
  {Glover}, {W{\"u}nsch}, {Klessen}, {Clark}, {Peters}, {Derigs}, \&
  {Baczynski}}]{Walch15}
{Walch} S. {et~al.}, 2015, \mnras, 454, 238

\bibitem[{{Walter} {et~al}\mbox{.}(2008){Walter}, {Brinks}, {de Blok},
  {Bigiel}, {Kennicutt}, {Thornley}, \& {Leroy}}]{Walter08a}
{Walter} F., {Brinks} E., {de Blok} W.~J.~G., {Bigiel} F., {Kennicutt}, Jr.
  R.~C., {Thornley} M.~D., {Leroy} A., 2008, \aj, 136, 2563

\end{thebibliography}

\appendix

\section{Analytic expression for the quadrupole potential} \label{appendix:quadrupole}
The quadrupole $\Phi_2$ used in Section \ref{sec:theory} is generated by the following density distribution:
\begin{equation}
 \rho_2(r,\phi,\theta) = \frac{A}{4 \pi G} \left(\frac{v_0 {\rm e}}{\rqu}\right)^2  \exp \left( - \frac{2 r}{\rqu} \right) \sin^2\phi\, \cos(2 \theta)  \;, \label{eq:rho2}
\end{equation}
where $A$ is a dimensionless parameter, $G$ is the gravitational constant, $v_0$ and $\rqu$ are parameters and $(r,\theta,\phi)$ are spherical coordinates where $r$ is the radial distance, $\phi$ is the polar angle and $\theta$ is the azimuthal angle. The plane $\phi=\pi/2$ coincides with the plane of the galaxy. In the main text we always assume $\phi=\pi/2$ since everything is two-dimensional. The values of the parameters used in the main text are $A=0.4$, $v_0 = 220 \kms$ and $\rqu = 1.5 \kpc$.

The potential corresponding to this density distribution can be found analytically. Since the density $\rho_2$ is proportional to the real part of the spherical harmonic $Y_2^2$, which is an eigenfunction of the Laplacian operator, the potential must be of the following form:
\begin{equation}
\Phi_2(r,\theta,\phi) = \Phi_2(r) \sin^2\phi\, \cos(2 \theta) .
\end{equation}
Taking the Laplacian of this equation and using the Poisson equation we find:
\begin{align}
\nabla^2 \Phi_2(r,\theta,\phi) 	& = \left[   \Phi_2'' + \frac{2}{r}\Phi_2'  - \frac{6}{r^2}\Phi_2  \right] \sin^2\phi\, \cos(2 \theta)  \\
						&= 4 \pi G \rho_2.
\end{align}
Hence $\Phi_2(r)$ must satisfy:
\begin{equation}\label{eq:diff}
 \Phi_2'' + \frac{2}{r}\Phi_2'  - \frac{6}{r^2}\Phi_2 = A \left(\frac{v_0 {\rm e}}{\rqu}\right)^2  \exp \left( - \frac{2 r}{\rqu} \right).
\end{equation}
We can reduce this to a dimensionless equation by defining the following dimensionless variables:
\begin{align}
x & = \frac{r}{\rqu} ,\\
F & = - \frac{\Phi_2}{ A \left( {v_0 {\rm e}} \right)^2}.
\end{align}
Equation \eqref{eq:diff} becomes:
\begin{equation}
F''(x) + \frac{ 2 F'(x)}{x} - \frac{ 6F(x)}{x^2}  = - \e^{-2x} .
\end{equation}
To find the potential corresponding to the density distribution \eqref{eq:rho2}, we need to solve this equation with boundary conditions
\begin{equation}
F(0) = F(\infty) = 0.
\end{equation}
The solution is 
\begin{equation}
F(x)= \frac{3 - e^{-2 x} \left(2 x^4+4 x^3+6 x^2+6 x+3\right) + 4 x^5 \operatorname{E_1}(2 x)}{20 x^3},
\end{equation}
where $\operatorname{E_1}(x)$ is the exponential integral function, a special function defined as
\begin{equation}
\operatorname{E_1}(x)\equiv \int_{x}^{\infty}\frac{e^{-t}}t\,dt.\,
\end{equation}

\section{Conservation of angular momentum for a fluid} \label{appendix:L}

Consider a fluid governed by the continuity equation and the compressible Navier-Stokes equation, which in conservative form are:
\begin{align}
\pa_t \rho + \pa_j \left( \rho v_j \right)				& = 0 \label{eq:continuity} \\
\pa_t (\rho v_i) +   \pa_j \left( \rho v_i v_j + \delta_{ij} P  \right) 			& = - \pa_j \left( \rho \sigma_{ij} \right) - \rho \pa_i \Phi \label{eq:ns}
\end{align}
where $\Phi$ is the external gravitational potential, the viscous tensor is given by \eqref{eq:sigmaij} and we use the convention that repeated indices means summation. From \eqref{eq:continuity} and \eqref{eq:ns} one can prove the following statement for the conservation of angular momentum:
\begin{equation} \label{eq:Lcons}
\pa_t (\rho L_i) + \pa_j ( \rho L_i v_j +  \epsilon_{ipm} x_p \delta_{mj} P) = \pa_j \left(\rho \epsilon_{ipm} x_p \sigma_{mj} \right) - \rho \epsilon_{ipm} x_p \pa_m \Phi
\end{equation}
where $L_i$ is the angular momentum per unit mass,
\begin{equation}
L_i = \epsilon_{ijk} x_j v_k.
\end{equation}
The first term on the right hand side of \eqref{eq:Lcons} represent the flux of angular momentum due to viscous forces, while the second represents the change in angular momentum due to gravitational forces. After some manipulation and using Eq. \eqref{eq:continuity}, Eq. \eqref{eq:Lcons} can be recast in the following form:
\begin{equation}
\frac{D \mathbf{L}}{Dt} =  \bm{\tau}_P + \bm{\tau}_\nu + \bm{\tau}_g
\end{equation}
where $D$ is the convective derivative
\begin{equation} \label{eq:Lcons2}
\frac{D}{Dt} = \pa_t + \mathbf{v} \cdot \nabla
\end{equation}
and the torques per unit mass due to pressure, viscous and gravitational forces are given by
\begin{align}
[\tau_P]_i & = - \frac{1}{\rho} \pa_j \left( \rho \epsilon_{ipm} x_p \delta_{mj} P \right), \\
[\tau_\nu]_i & = \frac{1}{\rho} \pa_j \left( \rho \epsilon_{ipm} x_p \sigma_{mj} \right), \label{eq:taunu2} \\
[\tau_g]_i & = - \epsilon_{ipm} x_p \pa_m \Phi.
\end{align}
The physical meaning of Eq. \eqref{eq:Lcons2} is that if we follow a fluid element, its angular momentum changes due to three contributions: pressure, viscous and gravitational torques. In two dimensions, Eq. \eqref{eq:taunu2} reduces to Eq. \eqref{eq:taunu}.

\end{document}